\documentclass[singlecolumn,amsmath,amssymb]{revtex4-1}
\usepackage{xcolor}
\usepackage{booktabs}
\usepackage{graphicx}
\usepackage{bm}
\usepackage{threeparttable}

\begin{document}
	

\title{Fibonacci-Modulation-Induced Multiple Topological Anderson Insulators}
\author{Ruijiang Ji}
\affiliation{Institute of Theoretical Physics and State Key Laboratory of Quantum Optics Technologies and Devices, Shanxi University, Taiyuan 030006, China}
\author{Zhihao Xu}
\email{xuzhihao@sxu.edu.cn}
\affiliation{Institute of Theoretical Physics and State Key Laboratory of Quantum Optics Technologies and Devices, Shanxi University, Taiyuan 030006, China}
\affiliation{Collaborative Innovation Center of Extreme Optics, Shanxi University, Taiyuan 030006, China}

\begin{abstract}
	\bf Topological Anderson insulators (TAIs) provide a mechanism for topological phase transitions in disordered systems and have implications for quantum material design. In this work, we investigate the emergence of multiple TAIs in a one-dimensional spin-orbit coupled (SOC) chain subject to Fibonacci modulation, which transforms a trivial band structure into a sequence of topologically nontrivial phases. This behavior is characterized by the appearance of zero-energy modes and changes in the $Z_2$ topological quantum number. As the SOC amplitude decreases, the number of TAI phases increases, a feature that is closely related to the fractal structure of the energy spectrum induced by Fibonacci modulation. In contrast to conventional TAI phases with fully localized eigenstates, the wave functions in the Fibonacci-modulated TAI phases display multifractal properties. This model can be experimentally realized using a Bose-Einstein condensate in a momentum-space lattice, where its topological transitions and multifractal features can be explored through quench dynamics.
\end{abstract}

\maketitle

\section*{Introduction}
Topological states of matter have been extensively studied in various systems, including condensed-matter materials \cite{MZH2010,zhangSC2011,DasT2016,Ryu2016,Schindler2024,SlagerRJ202401,SlagerRJ202402}, ultracold atoms \cite{Spielman2019,Bhattacharya2023}, superconducting circuits \cite{Lehnert2014,Tan2019,Sun2019,Tan2021}, electronic circuits \cite{zhangxd2021}, photonic lattices \cite{Lul2014,Ozawa2019,Vicencio2024,Qu2024,Rechtsman2024}, and mechanical systems \cite{Huber2016}. Topological insulators are characterized by gapless edge states that are immune to backscattering from weak disorders \cite{Raghu2008,Wangz2008,Wangz2009}, but these states typically cannot survive strong disorders due to Anderson localization \cite{Chong2017,Zhangs2017}. A reverse transition has been observed: disorder can induce the emergence of protected edge states and quantized transport even in a trivial band structure. This disorder-induced topological phase, known as the topological Anderson insulator (TAI) \cite{Shen2009,Groth2009,Wang2011,Shen2012,Titum2015,Orth2016,Jiang2019,Huang2022,Cui2022,Sun2024,Luzp2025}, has been theoretically predicted in systems such as the two-dimensional (2D) HgTe/CdTe quantum well \cite{Shen2009,Groth2009,Xie2009} and the 1D Su-Schrieffer-Heeger chain \cite{Schmiedt2014}, and experimentally realized in a variety of engineered systems, including 1D cold atomic wires \cite{Gadway2018}, 2D photonic waveguide arrays \cite{Szameit2018,LiuGG2020}, and photonic quantum walk \cite{XueP2022}. Both uncorrelated random \cite{Gadway2018,Dunlap1990,ZhangDW202202,Nava2023,Cinnirella2024,KangDW2024,GhoshAK2024} and quasiperiodic disorders in the form of a cosine modulation incommensurate with lattice spacing \cite{ZhuSL2021,ZhangDW2022,LuZP2022,Luzp2025,Padhan2024,XueQK2024} can give rise to the TAI phase. Expanding the range of disorder types that can induce the TAI phase, and exploring the associated topological and localization effects, remains a critical focus of ongoing research.

Quasicrystals provide platforms for studying exotic localization phenomena and topologically nontrivial behavior in physics \cite{Ouyang2024,DaiQ2023,YaoH2020,Modugno2014,Garreau2008,Rai2021,Haas2024,Kobialka2024,Sandberg2024}. Unlike randomly disordered systems, quasicrystals possess long-range order without periodicity \cite{Sbroscia2020,Jagannathan2021}. Recent advances in quasicrystal growth and synthetic engineering---achieved with atomic and photonic precision---have enabled the realization of quasiperiodic structures, among which the Fibonacci quasicrystal stands out as a prominent example. In a conventional Fibonacci-modulated chain, several features emerge that are characteristic of quasicrystalline order. The energy spectrum displays a self-similar, fractal structure known as a "Cantor set", which is neither entirely continuous nor discrete. This complex spectrum is accompanied by a highly nontrivial spatial structure of the eigenstates: single-particle wave functions are typically critical, meaning they are neither fully localized nor completely extended \cite{1983,Zilberberg2012,Liutong2025,Verbin2013,Mace2016,Mace2017,Bian2021,Jagannathan2021}. Such critical states arise directly from the underlying quasiperiodic order and result in multifractal properties for both the spectrum and the eigenfunctions. Furthermore, the system lacks translational symmetry but retains long-range order, giving rise to unique transport and localization phenomena that are distinct from those found in periodic or purely disordered systems. These features make the Fibonacci chain a paradigmatic platform for exploring the interplay between order, disorder, and topology in low-dimensional quantum systems \cite{Rai2021,Haas2024,Kobialka2024,Sandberg2024}. Synthetic Fibonacci chains have been realized in various platforms, including cold atoms\cite{Singh2015,Weld2024}, photonics \cite{Verbin2013,Kraus2012,Negro2003,Verbin2015}, polaritonics \cite{Tanese2014,Goblot2020}, and dielectric chains or circuits \cite{Reisner2023,Franca2024}, where the wave functions and topological properties of Fibonacci chains have been directly measured and investigated experimentally \cite{Dareau2017,Rai2021,LiuF2018,LiuF2019,XuDH2020}.

In this work, we report the emergence of multiple TAIs induced by Fibonacci modulation in a 1D spin-orbit coupled (SOC) chain, starting from a trivial band structure. This phenomenon is confirmed by the appearance of nontrivial zero modes, along with changes in the $\mathcal{Z}_2$ topological quantum number. The number of TAI phases increases as the SOC amplitude decreases. The fractal structure of the spectrum plays a crucial role in governing the emergence of these multiple TAI phases. Specifically, the self-similar splitting of the spectrum induced by the Fibonacci modulation facilitates the formation of multiple topologically nontrivial regions, which evolve as the SOC amplitude is varied. In contrast to conventional TAI phases, which exhibit fully localized eigenstates, the wave functions in the Fibonacci-modulated TAI phases display multifractal characteristics. Furthermore, our model can be experimentally realized in a Bose-Einstein condensate along the momentum lattice, where its topological transitions and multifractal behavior can be probed through quench dynamics.

\section*{Results and discussion}

\subsection*{Model and Hamiltonian}

We investigate a 1D SOC atomic chain subjected to a Fibonacci-modulated on-site potential. This system is described by the Hamiltonian
\begin{equation}
	H=\sum_{<i,j>}\Psi_i^{\dagger}\mathcal{R}_{ij}\Psi_{j}+\sum_{i} \Psi_i^{\dagger}\mathcal{U}_i\Psi_i, \label{eq1}
\end{equation}
where $\Psi_i^{\dagger}=(c^{\dagger}_{i\uparrow},c^{\dagger}_{i\downarrow})$, and $c^{\dagger}_{i\beta}$ creates a particle with spin $\beta=\{\uparrow,\downarrow\}$ at site $i$. The hopping matrix is given by $\mathcal{R}_{ij}=-t_{0}\sigma_z\pm it_{so}\sigma_y$, corresponding to hopping along the $\pm \hat{x}$ direction, respectively. Here, the diagonal elements of $\mathcal{R}$ describe spin-conserved hopping with amplitude $t_0$ (set as the energy unit, $t_0=1$), while the off-diagonal terms represent the spin-flip hopping between nearest neighbors with amplitude $t_{so}$. The second term in the Hamiltonian accounts for the on-site Fibonacci modulation \cite{Jagannathan2021}, expressed as
\begin{align}
	\mathcal{U}_i=\left[\lambda \mathrm{sgn}\left(\cos(2\pi\alpha i+\phi)-\cos(\pi\alpha)\right)+M\right]\sigma_z, \label{eq2}
\end{align}
where $M$ is a uniform Zeeman potential, $\lambda$ denotes the amplitude of quasiperiodic modulation, $\textup{sgn}[\cdots]$ is the sign function, and $\phi\in[0,2\pi)$ is an arbitrary phase used to generate different modulation configurations. A shift in the phase $\phi$ is equivalent to a translation in the site index and can therefore be eliminated by redefining the origin of the position. 
The modulation frequency $\alpha$ is chosen as an irrational Diophantine number, which can be constructed from a generalized $\kappa$-Fibonacci sequence defined by $F_{\nu+1}=F_{\nu-1}+\kappa F_{\nu}$ with $F_{0}=F_{1}=1$ \cite{Kohmoto1983}. The corresponding irrational number is given by $\alpha=\lim_{\nu\to\infty}F_{\nu-1}/F_{\nu}$, where $\kappa=1,2,3,\cdots$, yielding the so-called metallic mean family. For example, the golden mean ($\alpha_g=(\sqrt{5}-1)/2$) corresponds to $\kappa=1$, the silver mean ($\alpha_s=(\sqrt{2}-1)$) to $\kappa=2$, and the bronze mean ($\alpha_b=(\sqrt{13}-3)/2$) to $\kappa=3$, and so on. In our study, the system size is set as $L=F_{\nu}$ and the modulation frequency is approximated by the rational number $\alpha=F_{\nu-1}/F_{\nu}$. This generalized $\kappa$-Fibonacci sequence of quasiperiodic modulations can be constructed using various methods, including the substitution method, the cut-and-project method \cite{Ouyang2024,Jagannathan2021}, and the characteristic function method (see Supplementary Note 1 for details). Here, Eq.~\eqref{eq2} corresponds to the characteristic function method.

The Hamiltonian $H$ respecting the chiral symmetry defined by $\sigma_x$ in spin space governs a $\kappa$-Fibonacci-modulated SOC chain. Its counterpart with a constant uniform Zeeman potential realizes the 1D AIII class topological insulator \cite{LiuXJ2013}. In the crystalline case, where $\mathcal{U}_i=M\sigma_z$, the clean SOC model splits into two bands with eigenvalues $E_{\pm}=\pm\sqrt{\left(M-2\cos k\right)^2+4t_{so}^2\sin^2k}$, where $k$ is the wave vector. For $|M|>2$, regardless of whether $t_{so}\ne 0$, a topological phase transition occurs, indicated by the closing of the energy gap at $\left|M\right|=2$. This transition is further characterized by the vanishing of zero-energy edge modes and the disappearance of nontrivial topological invariants. In this work, we primarily focus on the case of $\kappa=1$, investigating the emergence of multiple TAI phases and multifractal features induced by the Fibonacci modulation. We also examine the cases with $\kappa=2$ and $3$ (see Supplementary Note 4 for details), where the presence of multiple TAI phases is similarly observed.

\subsection*{TAI and multiple TAIs}

\begin{figure*}[htbp]
	\centering
	\includegraphics[clip, width=0.46\columnwidth]{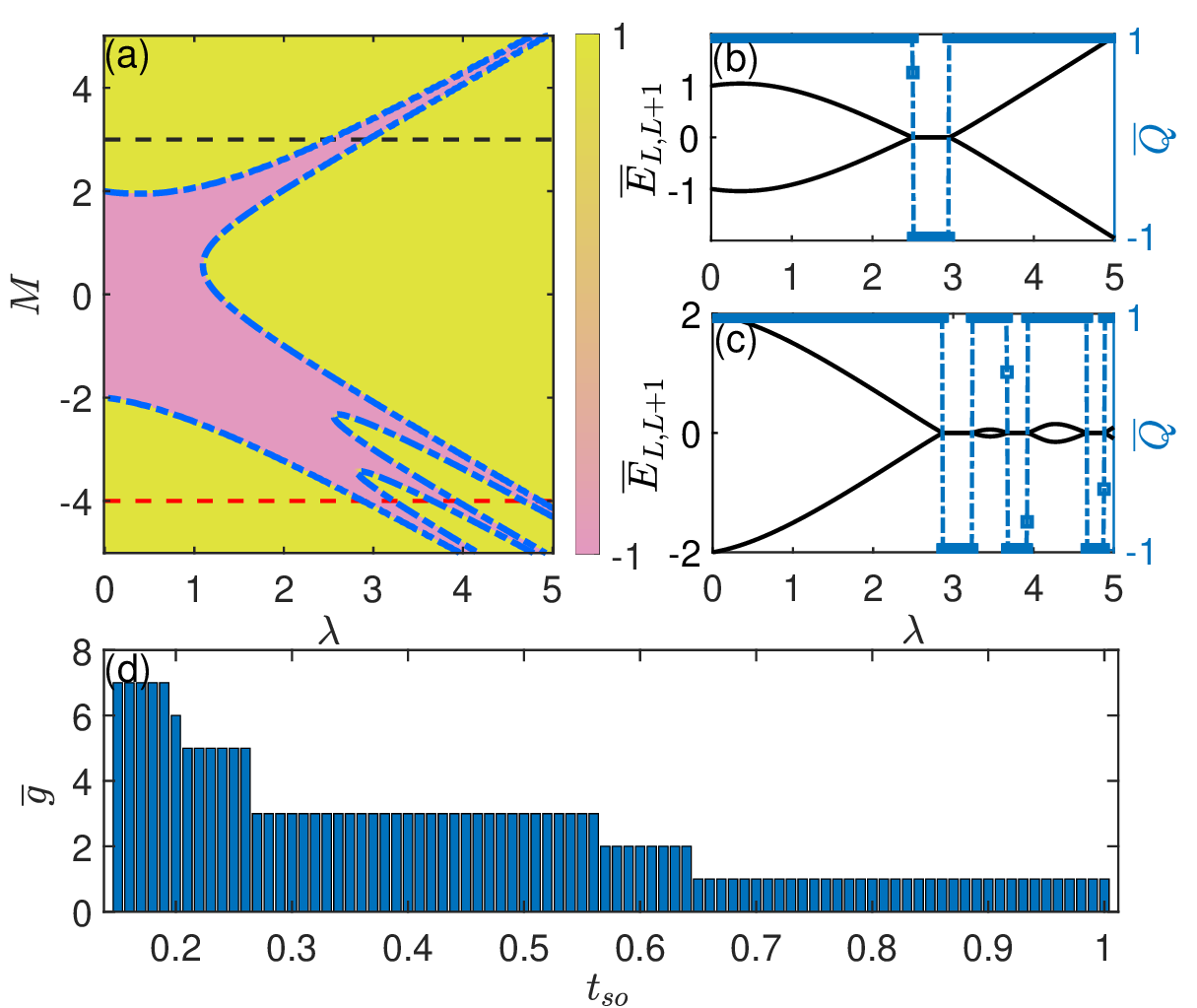}
	\caption{Topological transitions for different modulation and SOC strengths. (a) The disorder-averaged topological phase diagram with the spin-flip hopping strength $t_{so}=0.5$ as a function of the uniform Zeeman potential $M$ and the quasiperiodic modulation amplitude $\lambda$. The blue dashed lines correspond to the topological phase boundaries determined by Eq.(\ref{eq3}). The black (red) dashed line corresponds to the line of $M=3$ $(M=-4)$. Two disorder-averaged energies $\overline{E}_L$ and $\overline{E}_{L+1}$ at the center of the spectrum and the disorder-averaged topological number $\overline{Q}$ as a function of $\lambda$ under OBC with $t_{so}=0.5$ for (b) $M=3$ and (c) $M=-4$. (d) The number of times the TAI phases emerge, denoted as $\overline{g}$, as a function of $t_{so}$ with $M=-4$. Here, $\kappa=1$, and all the data are averaged by $N_c=50$ disorder realizations.}
	\label{Fig1}
\end{figure*}

Topologically protected edge states and quantized topological charges emerge in the context of TAI when sufficient disorder or incommensurate modulation is introduced to a trivial band structure. We employ the scattering-matrix method to compute the $\mathcal{Z}_2$ topological quantum number $Q$ in our case (see Methods), where $Q=1$ ($Q=-1$) corresponds to a topologically trivial (nontrival) phase. Figure \ref{Fig1}(a) shows the disorder-averaged topological phase diagram with $t_{so}=0.5$ and $\kappa=1$ as a function of $M$ and $\lambda$, obtained numerically averaging the disorder-averaged $\mathcal{Z}_2$ number $\overline{Q}$ over $N_c$ disorder realizations for different $\phi\in[0,2\pi)$. The phase boundaries can be determined by the condition:
\begin{equation}
	|\mathrm{Tr}(T)|-\mathrm{Det}(T)=1, \label{eq3}
\end{equation}
where $T$ is the total transfer matrix defined in a new basis $\tilde{\Psi}_i^{\dagger}=(c_{i\uparrow}^{\dagger}+c_{i\downarrow}^{\dagger},c_{i\uparrow}^{\dagger}-c_{i\downarrow}^{\dagger})$ for the zero modes (see Supplementary Note 2 for details). In the absence of $\lambda$, the system exhibits a topological phase transition at $|M|=2$, corresponding to a jump in $\overline{Q}$ from $-1$ to $1$ as $|M|$ increases. When the Fibonacci modulation ($\kappa=1$) is introduced to the nontrivial band structure with $|M|<2$, the topologically nontrivial phase remains robust against weak $\lambda$, but eventually transitions to a trivial one for sufficiently strong modulation. For $M>2$, the system undergoes a topological phase transition from trivial to nontrivial to trivial as $\lambda$ increases, indicating the emergence of the TAI. We further present two disorder-averaged energies, $\overline{E}_L$ and $\overline{E}_{L+1}$, at the center of the spectrum, as well as $\overline{Q}$ as a function of $\lambda$ under open boundary condition (OBC) with $\kappa=1$, $M=3$ and $t_{so}=0.5$ in Fig. \ref{Fig1}(b). When $\lambda$ lies in the range $[2.51,2.94]$, the topological number $\overline{Q}$ jumps from $1$ to $-1$, accompanied by the emergence of the zero edge modes. For $\lambda>2.94$, the topologically nontrivial phase vanishes. Our numerical results suggest that the Fibonacci modulation can induce the TAI.

\begin{figure*}[htbp]
	\centering
	\includegraphics[clip, width=0.46\columnwidth]{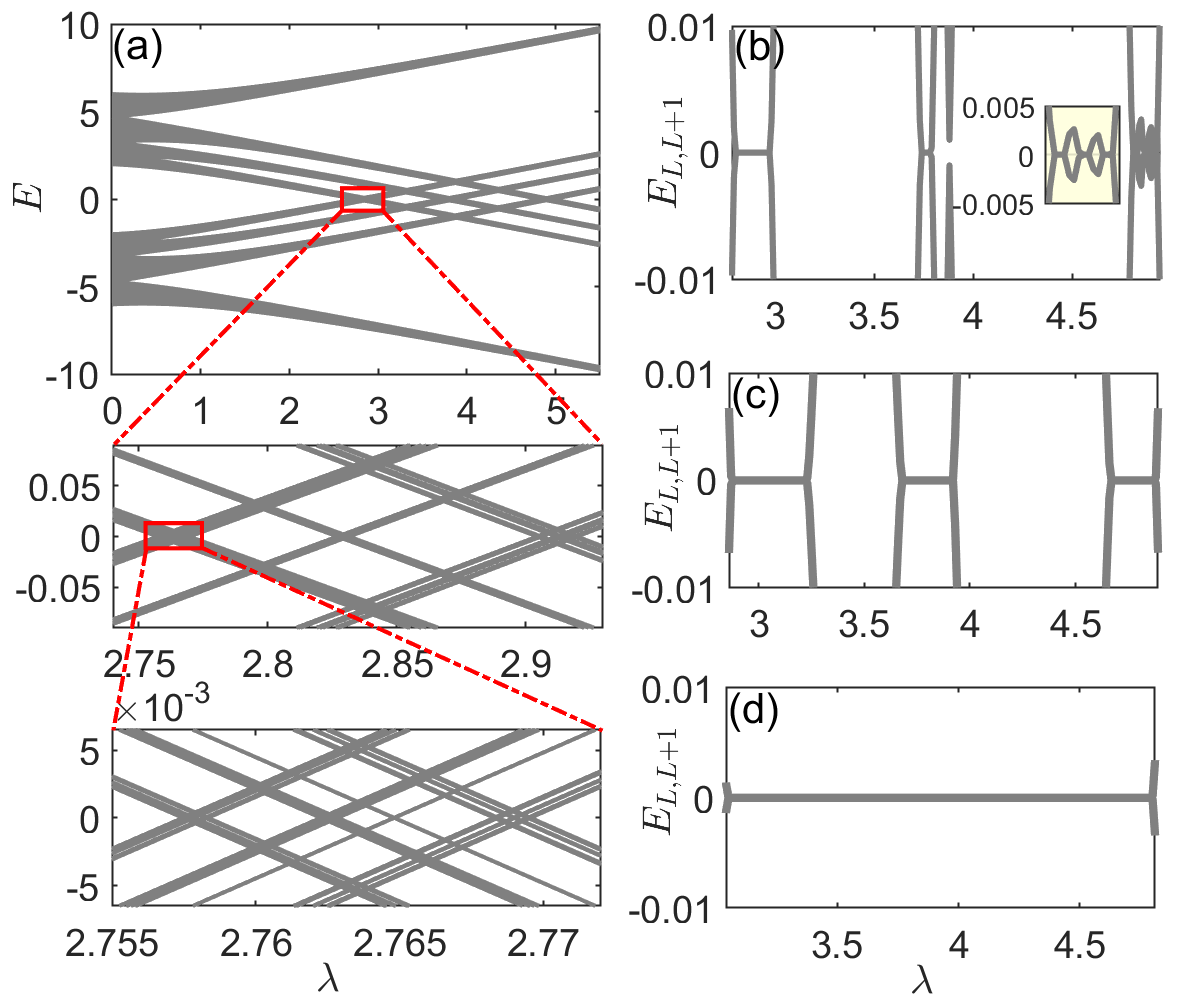}
	\caption{Energy spectra of the system with Fibonacci modulation for different SOC strengths. (a) Self-similarities of energy spectrum as a function of the quasiperiodic modulation amplitude $\lambda$ with the spin-flip hopping strength $t_{so}=0$. The insets in (a) show magnified views of the regions within the red boxes. Two central energy levels $E_{L}$ and $E_{L+1}$ as a function of $\lambda$ with (b) $t_{so}=0.25$, (c) $0.5$, and (d) $0.9$. The inset in (b) shows a magnified view of the original image. Here, the uniform Zeeman potential $M=-4$, $\kappa=1$, $\phi=0$, and $L=610$.}
	\label{Fig2}
\end{figure*}

As seen in Fig. \ref{Fig1}(a) for $M<-2$, the system undergoes multiple distinct transitions into the TAI phase, eventually returning to a trivial phase as $\lambda$ increases further. Figure \ref{Fig1}(c) shows the disorder-averaged energies of the two center modes, $\overline{E}_L$ and $\overline{E}_{L+1}$, as well as $\overline{Q}$ as a function of $\lambda$ with $\kappa=1$, $M=-4$ and $t_{so}=0.5$. It can be observed that, as the $\mathcal{Z}_2$ number $\overline{Q}$ transitions to a nontrivial value, corresponding zero modes emerge in the ranges $\lambda \in (2.87,3.22) \cup (3.67,3.94) \cup (4.66,4.88)$. The system undergoes three transitions from topologically trivial phases to nontrivial ones, indicating the emergence of the multiple TAI phases. For a given $M<-2$, the number of times TAI phases emerge, denoted by $g$, strongly depend on $t_{so}$. As shown in Fig. \ref{Fig1}(d) with $\kappa=1$ and $M=-4$, $\overline{g}$ gradually increases as $t_{so}$ decreases, for $N_c=50$ disorder realizations.

To uncover the underlying mechanism of this phenomenon, we analyze the energy spectrum of two decoupled Fibonacci-modulated chains ($\kappa=1$) without the SOC term ($t_{so}=0$) as described in Eq. (\ref{eq1}), shown in Fig. \ref{Fig2}(a) for $\phi=0$. For finite $\lambda$, the spectrum exhibits eight primary bands. Within the range $\lambda \in (2.76, 4.95)$, the middle six bands overlap, forming three clusters of band-crossing regions separated by two prominent band gaps near zero energy. Remarkably, each cluster undergoes successive splitting into smaller sub-clusters, creating a self-similar fractal structure that persists in the thermodynamic limit. This fractal nature manifests in the emergence of progressively smaller band gaps $\Delta_g$ at each level of splitting. When the SOC strength $t_{so}$ is introduced, these band gaps comparable in size to $t_{so}$ begin to close and reopen, signaling the formation of nontrivial zero-energy modes, as illustrated in Fig. \ref{Fig2}(b). This process unveils an infinite hierarchy of TAI phases in the small $t_{so}$ regime and in the thermodynamic limit, effectively creating a topologically fractal structure. As $t_{so}$ increases further, neighboring topologically nontrivial regions merge, reducing the number of distinct TAI phases. This transition, depicted in Figs. \ref{Fig2}(c) and \ref{Fig2}(d), signifies a transformation from a fractal topology to a more discrete structure, highlighting how the interplay between SOC and Fibonacci modulation reshapes the landscape of topological phases.

In Supplementary Note 3, we present an off-diagonal Fibonacci-modulated SOC chain ($\kappa=1$) and uncover the emergence of multiple TAI phases, whose occurrence strongly depends on $t_{so}$. The localization properties exhibit multifractal behavior, regardless of whether the system resides in a TAI phase or not for the off-diagonal Fibonacci-modulated SOC chain. These findings demonstrate that both diagonal and off-diagonal Fibonacci modulation in SOC chains can induce multiple TAI phases, offering insights into quasiperiodic topological systems. To study the modulation contribution to the emergence of TAIs, we investigate the cases with the modulation frequency chosen as other metallic mean numbers ($\kappa=2$ and $3$) in Supplementary Note 4. Our results show that, similar to the golden mean case, both the silver mean and bronze mean modulations lead to multiple reentrant TAI phases as the disorder strength increases. The topological phase diagrams for these cases reveal a sequence of transitions between trivial and topological phases, with the number of TAI regions and their evolution strongly affected by the modulation frequency and SOC strength. The energy spectra without SOC exhibit self-similar fractal structures. Introducing finite SOC causes band gaps to close and reopen, resulting in the emergence and eventual merging of TAI regions. These findings confirm that the appearance of multiple reentrant TAI phases is a universal and robust phenomenon for various metallic mean modulations, with the modulation frequency playing a key role in shaping the complexity and evolution of the topological phase diagram.

\subsection*{Localization properties}

\begin{figure*}[htbp]
	\centering
	\includegraphics[clip, width=1\columnwidth]{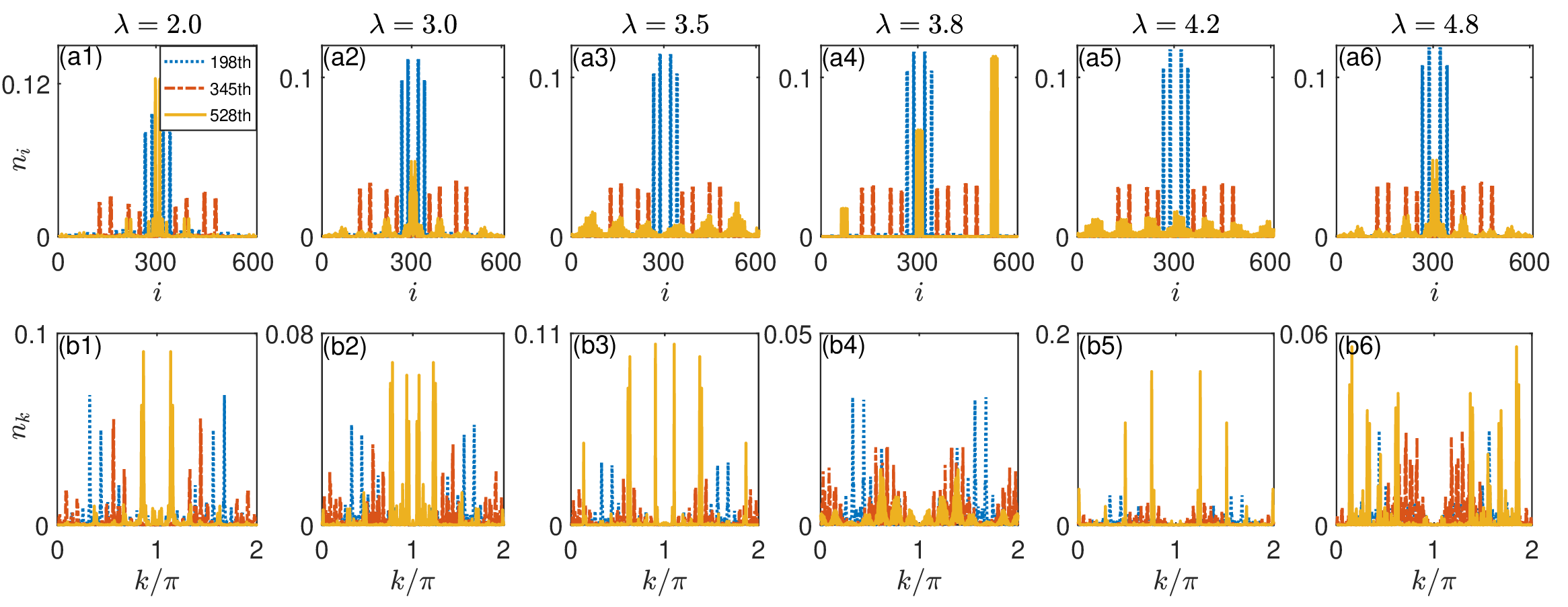}
	\caption{Density distributions in both real and momentum spaces. Density distributions of selected eigenstates for various values of the quasiperiodic modulation amplitude $\lambda$ in both real (a1)-(a6) and momentum (b1)-(b6) spaces with the spin-flip hopping strength $t_{so}=0.5$, the uniform Zeeman potential $M=-4$, $\kappa=1$, $\phi=0$, and $L=610$. From left to right, the columns correspond to $\lambda=2.0$, $3.0$, $3.5$, $3.8$, $4.2$, and $4.8$, where the system is localized in different topological regimes.}
	\label{Fig3}
\end{figure*}

A commonly held view is that in the TAI phase, all states are localized, exhibiting Anderson localization. However, in our case, the Fibonacci modulation fundamentally alters the localization properties. For any nonzero $\lambda$, nearly all the states display multifractal characteristics, regardless of the topological phase. Figure \ref{Fig3} shows the density distributions of several representative eigenstates in real and momentum spaces for the case of $\kappa=1$ with different $\lambda$ where the system is localized in both topologically trivial ($\lambda=2$, $3.5$, and $4.2$) and nontrivial ($\lambda=3$, $3.8$, and $4.8$) phases. Here, we select $n=198$, $345$, and $528$ as examples, where the eigenvalues $E_n$ are ordered in ascending order. The momentum-space density distribution \cite{Wang2020} is defined as $n_k=1/L\sum_{i,j}e^{-ik(i-j)}(\rho^{\uparrow}_{ij}+\rho^{\downarrow}_{ij})$, where $\rho^{(\beta)}_{ij}=\langle\Psi_n|c^{\dag}_{i\beta}c_{j\beta}|\Psi_n\rangle$ represents the single-particle density matrix of the $n$th eigenstate,  $|\Psi_n\rangle=\sum_{i,\beta} \psi_{i\beta}^{(n)}|i,\beta\rangle$, with the eigenvalue $E_n$. The real-space density distribution is given by $n_i=\rho_{ii}^{(\uparrow)}+\rho_{ii}^{(\downarrow)}$. In momentum space, a localized (extended) state exhibits an extended (localized) distribution \cite{Liutong2025,DaiQ2023}. Multifractal states, however, exhibit delocalized yet nonergodic behavior in both real and momentum spaces. The density distributions in both spaces confirm the multifractal nature of the states, which persists irrespective of whether the states are in a topologically nontrivial phase or not. 

\begin{figure*}[htbp]
	\centering
	\includegraphics[clip, width=1\columnwidth]{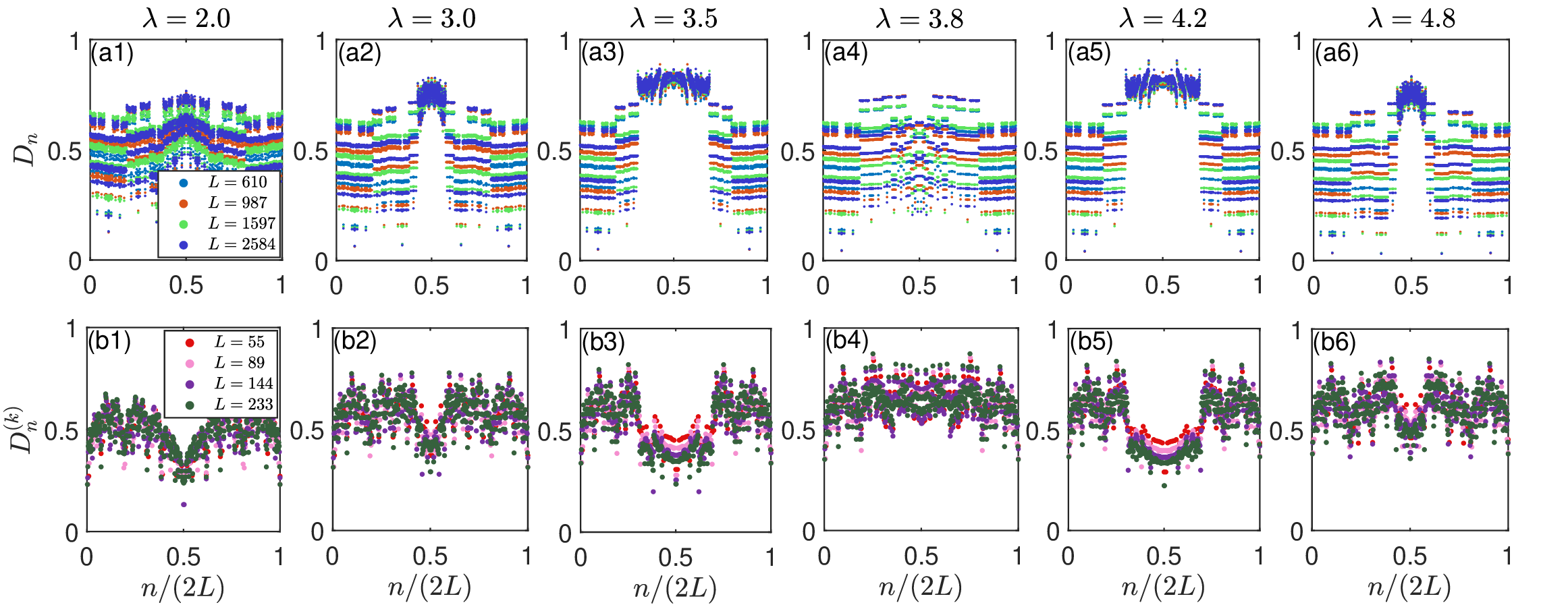}
	\caption{Fractal dimensions in both real and momentum spaces. Fractal dimensions for different system sizes under PBCs in both real (a1)-(a6) and momentum (b1)-(b6) spaces with the spin-flip hopping strength $t_{so}=0.5$, the uniform Zeeman potential $M=-4$, $\kappa=1$, and $\phi=0$. From left to right, the columns correspond to different quasiperiodic modulation amplitudes $\lambda=2.0$, $3.0$, $3.5$, $3.8$, $4.2$, and $4.8$, where the system is localized in different topological regimes.}
	\label{Fig4}
\end{figure*}

\begin{figure*}[htbp]
	\centering
	\includegraphics[clip, width=0.3\columnwidth]{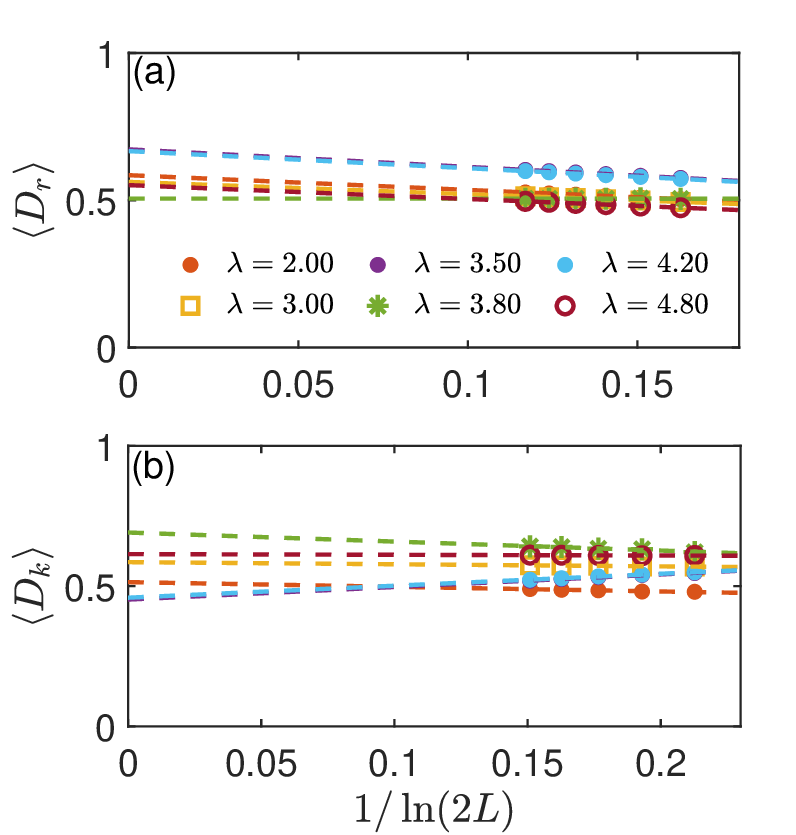}
	\caption{The scaling of the mean fractal dimensions in both real and momentum spaces. The scaling of the mean fractal dimensions in both (a) real and (b) momentum spaces for various values of the quasiperiodic modulation amplitude $\lambda$. The dashed lines correspond to the finite-size extrapolation of $\langle D_r \rangle$ ($\langle D_k \rangle$) as a function of $1/\ln(2L)$. Here, the spin-flip hopping strength $t_{so}=0.5$, the uniform Zeeman potential $M=-4$, $\kappa=1$, and $\phi=0$.}
	\label{Fig5}
\end{figure*}

By contrasting the fractal dimensions for each eigenstate at different system sizes in both real and momentum spaces, we can clearly characterize the localization properties of our system across distinct topological phases as the modulation amplitude increases. This analysis is illustrated in Fig. \ref{Fig4} for the case of $\kappa=1$, $t_{so}=0.5$, $M=-4$, and $\phi=0$, under periodic boundary conditions (PBCs) for various values of $\lambda$. The fractal dimension of $n$th eigenstate in real space is defined as $D_n=-\lim_{L\to\infty}\ln{(\mathrm{IPR}_n)}/\ln{(2L)}$, where the inverse participation ratio (IPR) is given by $\mathrm{IPR}_n=\sum_{i,\beta}|\psi_{i\beta}^{(n)}|^4$. In the thermodynamic limit, $D_n\to 0$ for localized states and $D_n\to 1$ for extended states. When $0<D_n<1$, the eigenstate is considered multifractal. Multifractality can also be detected by analyzing the fractal dimension in momentum space, defined as  $D^{(k)}_n=-\lim_{L\to\infty}\ln{(\mathrm{IPR}^{(k)}_n)}/\ln{(2L)}$, where  $\mathrm{IPR}^{(k)}_n=\sum_{l}n^2_{k_l}$ with $k_l=2\pi l/L$ $(l=1,...,L)$. For extended (localized) states in the real space, $D^{(k)}_n$ extrapolates to $0$ $(1)$ , whereas in the multifractal regime, $D^{(k)}_n$ takes values significantly different from both $0$ and $1$. As shown in Fig. \ref{Fig4}, for both topologically trivial and nontrivial phases, the fractal dimensions of nearly all states in both real and momentum spaces remain system-size independent, and deviate from the limiting values of $0$ and $1$. To further explore the thermodynamic limit, we analyze the scaling behavior of the mean fractal dimension in the real space, $\langle D_r\rangle=1/(2L)\sum_{n}D_n$, and in the momentum space, $\langle D_k \rangle=1/(2L)\sum_{n}D^{(k)}_n$,  as shown in Fig. \ref{Fig5}. In the thermodynamic limit, both $\langle D_r\rangle$ and $\langle D_k\rangle$ approach values distinct from both $0$ and $1$ in all phases. Specially, we fit the scaling behavior for the both mean fractal dimensions: $\langle D_r \rangle=a_r/\ln(2L)+\langle D_r^{\infty} \rangle$ and $\langle D_k \rangle=a_k/\ln(2L)+\langle D_k^{\infty} \rangle$, respectively. Table \ref{table1} summarizes the extrapolated mean fractal dimensions $\langle D_r^{\infty} \rangle$ and $\langle D_k^{\infty} \rangle$ for selected values of $\lambda$ in different topological phases, with $\kappa=1$, $t_{so}=0.5$, $M=-4$, and $\phi=0$. Our results indicate that eigenstates of the Fibonacci modulated chain exhibit robust multifractal characteristics across different topological phases, and the emergence of the TAI phase does not alter the underlying localization properties of the system.

\begin{table}[htbp]
	\centering
	\caption{The extrapolated mean fractal dimensions $\langle D_r^{\infty} \rangle$ and $\langle D_k^{\infty} \rangle$ in different topological phases.}
	\label{table1}
	
	\begin{tabular}{lcccccc}
		\toprule[2pt]
		$\lambda$ & ~2.0 & ~3.0 & ~3.5 & ~3.8 & ~4.2 & ~4.8 \\
		\midrule[1pt]
		$\langle D_r^{\infty} \rangle$ & ~0.5850 & ~0.5621 & ~0.6727 & ~0.5054 & ~0.6667 & ~0.5513     \\
		$\langle D_k^{\infty} \rangle$ & ~0.5144 & ~0.5857 & ~0.4522 & ~0.6912 & ~0.4589 & ~0.6144   \\
		\bottomrule[1pt]
	\end{tabular}
	\begin{tablenotes}
		\item[a] The system is localized in both topologically trivial (the quasiperiodic modulation amplitude $\lambda=2.0$, $3.5$, and $4.2$) and nontrivial ($\lambda=3.0$, $3.8$, and $4.8$) phases. Here, the spin-flip hopping strength $t_{so}=0.5$, the uniform Zeeman potential $M=-4$, $\kappa=1$, and $\phi=0$.
	\end{tablenotes}
\end{table}

\subsection*{Dynamical detection}

\begin{figure*}[htbp]
	\centering
	\includegraphics[clip, width=0.46\columnwidth]{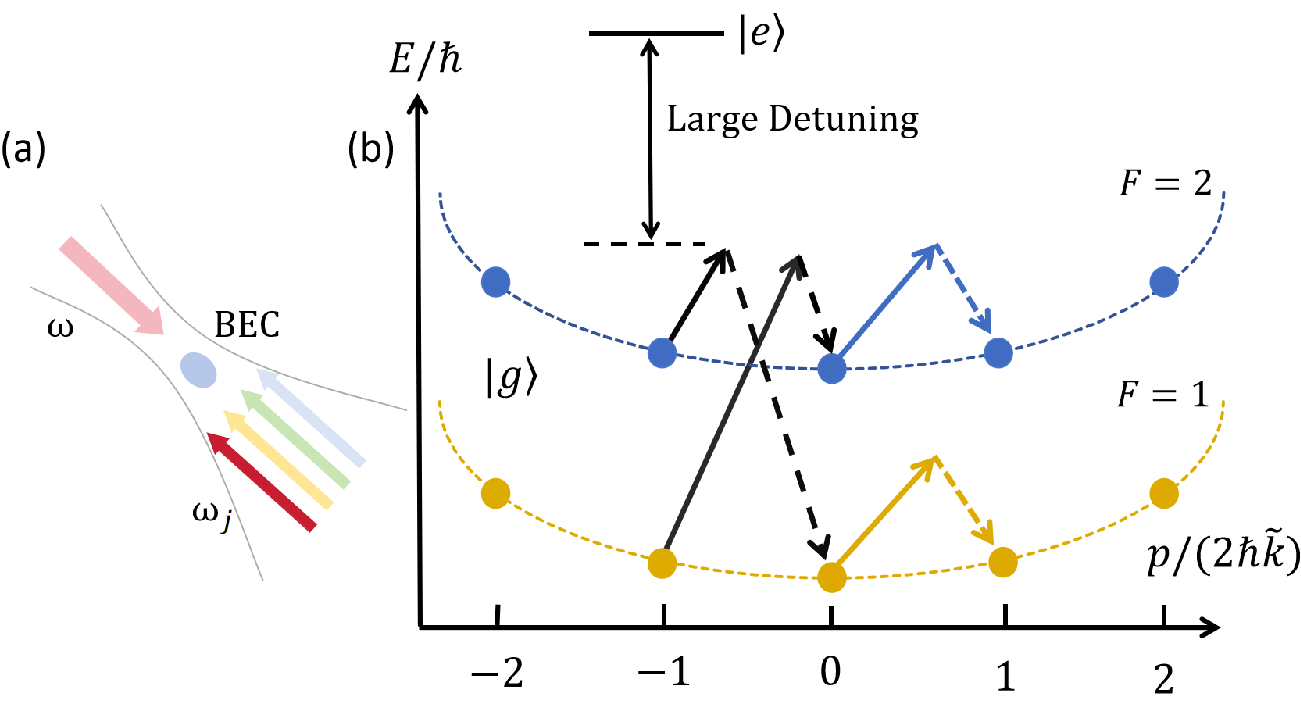}
	\caption{Illustration of experimental scheme. (a) A quasi-1D BEC is illuminated by a pair of counterpropagating laser beams: one with a fixed frequency $\omega$ and the other containing multifrequency components $\omega_j$. (b)  The lasers, far detuned from the atomic transition, drive a series of engineered two-photon Bragg transitions. These transitions couple different momentum states within the same ground-state hyperfine manifold (blue and yellow arrows) or across different ground-state hyperfine manifolds (black arrows), with an increment of $2\hbar \tilde{k}$, where $\tilde{k}=2\pi/\tilde{\lambda}$ with $\tilde{\lambda}$ being the wave length of the lasers. The solid and dashed arrows represent the processes of photon absorption and emission, respectively.}
	\label{Fig6}
\end{figure*}

Experimentally, this Fibonacci-modulated SOC chain can be realized along the momentum lattice \cite{Gadway2015,Gadway2016,Gadway2017,Gadway2018,Gadway201802,WangYF2022,LiYQ2022,YanBo2022,YanBo202202,LiYQ2023,YanBo2024,Gadway2024} in a cold atom of Bose-Einstein condensate, 
as illustrated in Fig. \ref{Fig6}. In this setup, the spin and lattice-site degrees of freedom are encoded in the two ground-state hyperfine manifolds of selected atoms, such as $^{87}$Rb \cite{YanBo2024}. Specifically, the hyperfine state $F=1$  represents spin-down, while $F=2$ represents spin-up. A pair of counter-propagating laser beams is employed: the incoming laser beam has a fixed frequency, while the acousto-optic modulator in the reflected laser beam generate multi-frequency components. These laser beams drive a series of two-photon Bragg transitions between adjacent discrete momentum states. Transitions within the same hyperfine manifold enable nearest-neighbor couplings with the same spin, corresponding to the diagonal terms of $\mathcal{R}$. In contrast, transitions between different hyperfine manifolds facilitate nearest-neighbor couplings with opposite spins, corresponding to the off-diagonal terms of $\mathcal{R}$. By precisely adjusting the amplitude of each multi-frequency component, all couplings can be individually controlled. Additionally, the on-site potentials $\mathcal{U}_i$ can be tuned via the detuning of the two-photon Bragg transitions.

\begin{figure*}[htbp]
	\centering
	\includegraphics[clip, width=0.5\columnwidth]{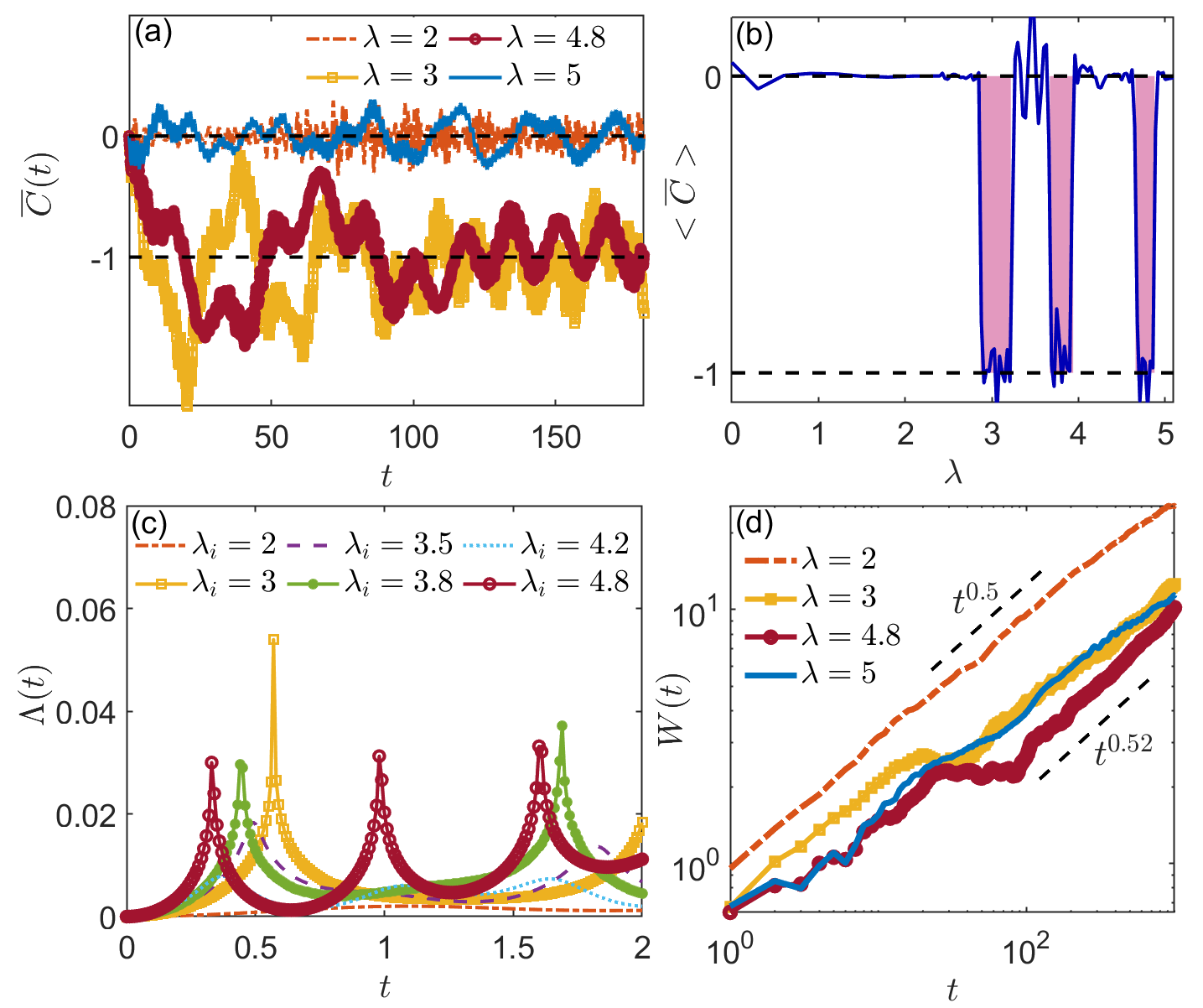}
	\caption{Dynamical detections of topological and localization properties. (a) The disorder-averaged expectation value of the chiral displacement operator $\overline{C}(t)$ versus time $t$ for different quasiperiodic modulation amplitudes $\lambda$ with $N_c=50$ disorder realizations. (b) The time average of the chiral displacement operator $\langle\overline{C}\rangle$ as a function of the quasiperiodic modulation amplitude $\lambda$ with $N_c=200$ disorder realizations. The pink area represents that the system is localized in topologically nontrivial phase. (c) The evolution of the dynamical free energy $\Lambda(t)$ with $\lambda_f=0$, calculated for the $(L+1)$th eigenstate at different values of $\lambda_i$ as the initial states. (d) The evolution of the mean-square displacement $W(t)$ for different $\lambda$ with $\phi=0$. The black dashed line indicates the linear fit, used to quantify the slope magnitude. Here, the spin-flip hopping strength $t_{so}=0.5$, the uniform Zeeman potential $M=-4$, and $L=144$.}
	\label{Fig7}
\end{figure*}

Topological properties and localization features of the system can be detected by the quench dynamics. To characterize the topology of our model, we monitor the dynamical response of the system to a sudden quench. Specifically, we measure the mean chiral displacement \cite{Gadway2018,Luzp2025,XueQK2024,Massignan2017,ZhangDW202202} and the Loschmidt echo (LE) \cite{ZhouBZ2021,ZhouBZ2019,Nehra2024,Langari2021,ChenB2024,HuZX2024}. We define the disorder-averaged expectation value of the chiral displacement operator as $\overline{C}(t) = 2 \langle\Psi(t)\left|\Gamma X\right|\Psi(t)\rangle$, where $\Gamma$ is the chiral operator, $X$ is the unit cell operator, and $|\Psi(t)\rangle=e^{iHt}|\Psi(0)\rangle$ is the time-evolved wave function, with the initial state $|\Psi(0)\rangle = c_{j_0\uparrow}^{\dagger} |0\rangle$ and $j_0$ being the position of the central bulk lattice site. The dynamics of $\overline{C}(t)$ generally exhibit transient, oscillatory behavior, as shown in Fig. \ref{Fig7}(a) for different $\lambda$ with $M=-4$ and $t_{so}=0.5$. In long time limit, $\bar{C}(t)$ converges to $0$ for $\lambda$ values corresponding to the topologically trivial phase and to $-1$ for those corresponding to the nontrivial phase. To eliminate the oscillation, we compute the time average $\langle\overline{C}\rangle$, which converges to $-1$ for a topologically nontrivial phase and to $0$ for a trivial one. Figure \ref{Fig7}(b) shows $\langle\overline{C}\rangle$ as the function of the modulation amplitude $\lambda$ with $M=-4$ and $t_{so}=0.5$. The value of $\langle\overline{C}\rangle$ starts at $0$, then jumps to a nontrivial value three times, before eventually returning to $0$. These results demonstrate that the mean chiral displacement is a sensitive probe of the multiple TAI phases in our model, effectively capturing the topological transitions driven by the modulation amplitude $\lambda$.

The LE, defined as $\mathcal{L}(t)=|\langle \Psi(0)| e^{-iH(\lambda_f)t} |\Psi(0) \rangle|^2$, is a powerful tool for analyzing nonequilibrium dynamics. It exhibits a series of zero points at specific time intervals when the initial Hamiltonian [$H(\lambda_i)$] and the post-quench Hamiltonian [$H(\lambda_f)$] describe systems that are localized in different phases. This behavior is a hallmark of dynamical quantum phase transitions (DQPTs), which have been observed in various systems\cite{Heyl2013,Heyl2014,Heyl2015,Jafari2017,Heyl2018,LiJun2020,Bandyopadhyay2021}. Pioneering studies have established a connection between DQPTs and emergent topological transitions. In particular, dynamical topological order parameters can change their integer values when DQPTs occur, providing a means to dynamically track the topological changes during the quench. The appearance of zero points in the LE signals the nonanalytic behavior in the dynamical free energy $\Lambda(t) =-1/(2L)\ln\mathcal{L}(t)$. Figure \ref{Fig7}(c) shows the evolution of $\Lambda(t)$ with $\lambda_f=0$, calculated for the $(L+1)$th eigenstate at different values of $\lambda_i$ for $M=-4$ and $t_{so}=0.5$ as the initial states. For the topologically trivial Hamiltonian $H(\lambda_f)$, when the initial Hamiltonian $H(\lambda_i)$ is in the topologically nontrivial phase ($\lambda_i=3$, $3.8$, and $4.8$), a serial of divergence points of $\Lambda(t)$--corresponding to exact zeros of LE--emerge along the time axis. In contrast, for $\lambda_i$ values within the topologically trivial phase ($\lambda_i=2$, $3.5$, and $4.2$), the nonanalytic behavior disappears, indicating that both the initial and the post-quench Hamiltonians are localized in the same topological phase. These results suggest that the LE method not only captures the occurrence of DQPTs but also provides a powerful tool to probe the emergence of multiple TAI phases in our system.

To dynamically detect the localization properties of our system, we employ the mean-square displacement \cite{Luzp2025,Rosch2012,BlochI2018,Wang2020} defined as
\begin{equation}
	W(t)=\left[\sum_{j,\beta=\uparrow,\downarrow} (j-j_0)^2\langle \Psi(t)| c^{\dagger}_{j,\beta}c_{j,\beta}|\Psi(t)\rangle\right]^{1/2}, \label{eq5}
\end{equation}
where the value of $W(t)$ grows in a power-law form of time, $W(t)\propto t^{\kappa}$, during the expansion process. Figure \ref{Fig6}(d) shows the time evolution of $W(t)$ for different values of $\lambda$ in various topological phases, with the initial state $|\Psi(0)\rangle = c_{j_0\uparrow}^{\dagger} |0\rangle$ and $j_0=L/2+1$ for even $L$. As seen in Fig. \ref{Fig7}(d), $W(t)$ exhibits subdiffusive behavior with $\kappa\approx 0.5$, which corresponds to the multifractal phase. Our results shows that the localization properties in different topological phases remain unchanged, and $W(t)$ serves as an effective tool to detect these properties.

\section*{Conclusion}

We have demonstrated that multiple TAI phases can be induced by Fibonacci modulation in a SOC chain. Unlike conventional TAI phases, where the system exhibits full localization, our results show that the system retains multifractal features, regardless of whether it is in the TAI phase or not. Crucially, the emergence of multiple TAI phases is closely tied to the fractal structure of the spectrum induced by the Fibonacci modulation. The self-similar splitting of the bands at each scale enables the formation of multiple TAI phases, resulting in a topologically nontrivial structure that evolves as the SOC amplitude is varied. This model can be experimentally realized along the momentum lattice in a cold atomic system, and its properties can be tested through dynamical evolution experiments.

\section*{Methods\label{methods}}

To determine the topological properties of the quasiperiodic-modulated SOC chain, we utilize the scattering matrix $S$, which relates the incoming and outgoing wave amplitudes at the Fermi level \cite{Zhang2016}:
\begin{equation}
	S=\begin{pmatrix}
		\tilde{R}_{\leftarrow} & \tilde{T}_{\leftarrow} \\
		\tilde{T}_{\rightarrow} & \tilde{R}_{\rightarrow}\\
	\end{pmatrix},
\end{equation}
where $\tilde{R}_{\leftarrow}$ and $\tilde{R}_{\rightarrow}$ are $2\times 2$ reflection matrices at the left and right ends of the chain, respectively, while  $\tilde{T}_{\leftarrow}$ and $\tilde{T}_{\rightarrow}$ are the corresponding transmission matrices. The $\mathcal{Z}_2$ topological quantum number $Q$ is defined as:
\begin{equation}
	Q=\mathrm{sgn}\left[\mathrm{Det}\left(\tilde{R}_{\leftarrow}\right)\right]=\mathrm{sgn}\left[\mathrm{Det}\left(\tilde{R}_{\rightarrow}\right)\right],
	\label{eq10}
\end{equation}
where $\mathrm{sgn}[\cdots]$ denotes the sign function. A value of $Q=1$ indicates a topologically trivial phase, while $Q=-1$ signifies a topologically nontrivial phase. To account for the effects of disorder, we define the disorder-averaged topological invariant $\overline{Q}$, calculated over $N_c$ disorder realizations. In our analysis, $N_c$ configurations are sampled for different values of $\phi\in[0,2\pi]$.

The scattering matrix can be derived using the transfer-matrix scheme. Based on the Hamiltonian given in Eq. (1) of the main text, the zero-energy Schr\"{o}dinger equation leads to the following recursive relation:
\begin{equation}
	\begin{pmatrix}
		\mathcal{R}_{i+1,i}^{\dagger}\Psi_i \\
		\Psi_{i+1}\end{pmatrix}=\tilde{B}_i \begin{pmatrix}
		\mathcal{R}_{i,i-1}^{\dagger}\Psi_{i-1}\\ \Psi_{i} \end{pmatrix}
\end{equation}
where
\begin{equation}
	\tilde{B}_i=\begin{pmatrix}
		0 & \mathcal{R}_{i+1,i}^{\dagger} \\
		-\mathcal{R}_{i,i+1}^{-1} & -\mathcal{R}_{i,i+1}^{-1}\mathcal{U}_i\\
	\end{pmatrix}.
\end{equation}
The relation indicates that wave amplitudes at the two ends of the chain ($i=1$ and $i=N$) are connected by the total transfer matrix $\tilde{B}=\tilde{B}_L\tilde{B}_{L-1}\ldots\tilde{B}_1$. To separate the right-moving and left-moving waves into the upper and lower four components, we transform the transfer matrix using the basis rotation:
\begin{equation}
	B_i=U^{\dagger}\tilde{B}_iU,
\end{equation}
where
\begin{equation}
	U=\frac{1}{\sqrt{2}}\begin{pmatrix}
		I & I \\
		iI & -iI \end{pmatrix},
\end{equation}
and $I$ is a $2\times2$ identity matrix. In this base, the reflection $(\tilde{R}_{\leftarrow},\tilde{R}_{\rightarrow})$ and transmission $(\tilde{T}_{\leftarrow},\tilde{T}_{\rightarrow})$ matrices can be determined via the relations:
\begin{equation}
	\begin{pmatrix}
		\tilde{T}_{\rightarrow}  \\
		0 \end{pmatrix}=B\begin{pmatrix}
		I\\
		\tilde{R}_{\leftarrow} \end{pmatrix}, \quad
	\begin{pmatrix}
		\tilde{R}_{\rightarrow}  \\
		I \end{pmatrix}=B\begin{pmatrix}
		0\\
		\tilde{T}_{\leftarrow} \end{pmatrix},
\end{equation}
where $B=B_LB_{L-1}\ldots B_1$.

\bibliography{References}

\begin{thebibliography}{10}
    \expandafter\ifx\csname url\endcsname\relax
	  \def\url#1{\texttt{#1}}\fi
	\expandafter\ifx\csname urlprefix\endcsname\relax\def\urlprefix{URL }\fi
	\providecommand{\bibinfo}[2]{#2}
	\providecommand{\eprint}[2][]{\url{#2}}
    \bibitem{MZH2010}
    \bibinfo{author}{Hasan, M. Z.} \& \bibinfo{author}{Kane, C. L.}
    \newblock \bibinfo{title}{Colloquium: Topological insulators}.
    \newblock \emph{\bibinfo{journal}{Rev. Mod. Phys.}}
    \textbf{\bibinfo{volume}{82}}, \bibinfo{pages}{3045}
	(\bibinfo{year}{2010}).
	
    \bibitem{zhangSC2011}
    \bibinfo{author}{Qi, X.-L.} \& 
    \bibinfo{author}{Zhang, S.-C.}
    \newblock \bibinfo{title}{Topological insulators and superconductors}.
    \newblock \emph{\bibinfo{journal}{Rev. Mod. Phys.}}
    \textbf{\bibinfo{volume}{83}}, \bibinfo{pages}{1057}
    (\bibinfo{year}{2011}). 
    \bibitem{DasT2016}
    \bibinfo{author}{Bansil, A.}, 
    \bibinfo{author}{Lin, H.} \& 
    \bibinfo{author}{Das, T.}
    \newblock \bibinfo{title}{Colloquium: Topological band theory}.
    \newblock \emph{\bibinfo{journal}{Rev. Mod. Phys.}}
    \textbf{\bibinfo{volume}{88}}, \bibinfo{pages}{021004}
    (\bibinfo{year}{2016}).
    \bibitem{Ryu2016}
    \bibinfo{author}{Chiu, C.-K.}, 
    \bibinfo{author}{Teo, J. C. Y.},
    \bibinfo{author}{Schnyder, A. P.} \& 
    \bibinfo{author}{Ryu, S.}
    \newblock \bibinfo{title}{Classification of topological quantum matter with symmetries}.
    \newblock \emph{\bibinfo{journal}{Rev. Mod. Phys.}}
    \textbf{\bibinfo{volume}{88}}, \bibinfo{pages}{035005}
    (\bibinfo{year}{2016}).
    \bibitem{Schindler2024}
    \bibinfo{author}{Davenport, H.}, 
    \bibinfo{author}{Knolle, J.} \& 
    \bibinfo{author}{Schindler, F.}
    \newblock \bibinfo{title}{Interaction-induced crystalline topology of excitons}.
    \newblock \emph{\bibinfo{journal}{Phys. Rev. Lett.}}
    \textbf{\bibinfo{volume}{133}}, \bibinfo{pages}{176601}
    (\bibinfo{year}{2024}).
    \bibitem{SlagerRJ202401}
    \bibinfo{author}{Thompson, J. J. P.}, 
    \bibinfo{author}{Jankowski, W. J.},
    \bibinfo{author}{Slager, R.-J.} \& 
    \bibinfo{author}{Monserrat, B.} 
    \newblock \bibinfo{title}{Topologically-enhanced exciton transport} \newblock \eprint{arXiv:2410.00967v1}
    (\bibinfo{year}{2024}).
    \bibitem{SlagerRJ202402}
    \bibinfo{author}{Jankowski, W. J.}, 
    \bibinfo{author}{Thompson, J. J. P.},
    \bibinfo{author}{Monserrat, B.} \& 
    \bibinfo{author}{Slager, R.-J.} 
    \newblock \bibinfo{title}{Excitonic topology and quantum geometry in organic semiconductors}.
    \newblock \emph{\bibinfo{journal}{Nat. Commun.}}
    \textbf{\bibinfo{volume}{16}}, \bibinfo{pages}{4661}
    (\bibinfo{year}{2025}).

    \bibitem{Spielman2019}
    \bibinfo{author}{Cooper, N. R.}, 
    \bibinfo{author}{Dalibard, J.} \& 
    \bibinfo{author}{Spielman, I. B.}
    \newblock \bibinfo{title}{Topological bands for ultracold atoms}.
    \newblock \emph{\bibinfo{journal}{Rev. Mod. Phys.}}
    \textbf{\bibinfo{volume}{91}}, \bibinfo{pages}{015005}
    (\bibinfo{year}{2019}).
    \bibitem{Bhattacharya2023}
    \bibinfo{author}{Jalali-mola, Z.},
    \bibinfo{author}{Grass, T.}, 
    \bibinfo{author}{Kasper, V.},
    \bibinfo{author}{Lewenstein, M.} \& 
    \bibinfo{author}{Bhattacharya, U.}
    \newblock \bibinfo{title}{Topological Bogoliubov quasiparticles from Bose-Einstein condensate in a flat band system}.
    \newblock \emph{\bibinfo{journal}{Phys. Rev. Lett.}}
    \textbf{\bibinfo{volume}{131}}, \bibinfo{pages}{226601}
    (\bibinfo{year}{2023}). 
 
    \bibitem{Lehnert2014}
    \bibinfo{author}{Schroer, M. D.} \emph{et~al.}
    \newblock \bibinfo{title}{Measuring a topological transition in an artificial spin-1/2 system}.
    \newblock \emph{\bibinfo{journal}{Phys. Rev. Lett.}}
    \textbf{\bibinfo{volume}{113}}, \bibinfo{pages}{050402}
    (\bibinfo{year}{2014}).
    \bibitem{Tan2019}
    \bibinfo{author}{Tan, X.} \emph{et~al.}
    \newblock \bibinfo{title}{Experimental measurement of the quantum metric tensor and related topological phase transition with a superconducting qubit}.
    \newblock \emph{\bibinfo{journal}{Phys. Rev. Lett.}}
    \textbf{\bibinfo{volume}{122}}, \bibinfo{pages}{210401}
    (\bibinfo{year}{2019}).
    \bibitem{Sun2019}
    \bibinfo{author}{Cai, W.} \emph{et~al.}
    \newblock \bibinfo{title}{Observation of topological magnon insulator states in a superconducting circuit}.
    \newblock \emph{\bibinfo{journal}{Phys. Rev. Lett.}}
    \textbf{\bibinfo{volume}{123}}, \bibinfo{pages}{080501}
    (\bibinfo{year}{2019}).
    \bibitem{Tan2021}
    \bibinfo{author}{Tan, X.} \emph{et~al.}
    \newblock \bibinfo{title}{Experimental observation of tensor monopoles with a superconducting qudit}.
    \newblock \emph{\bibinfo{journal}{Phys. Rev. Lett.}}
    \textbf{\bibinfo{volume}{126}}, \bibinfo{pages}{017702}
    (\bibinfo{year}{2021}). 
    \bibitem{zhangxd2021}
    \bibinfo{author}{Zhang, W.} \emph{et~al.}
    \newblock \bibinfo{title}{Experimental observation of higher-order topological Anderson insulators}.
    \newblock \emph{\bibinfo{journal}{Phys. Rev. Lett.}}
    \textbf{\bibinfo{volume}{126}}, \bibinfo{pages}{146802}
    (\bibinfo{year}{2021}).
    \bibitem{Lul2014}
    \bibinfo{author}{Lu, L.},
    \bibinfo{author}{Joannopoulos, J. D.} \& 
    \bibinfo{author}{Solja\v{c}i\'{c}, M.}
    \newblock \bibinfo{title}{Topological photonics}.
    \newblock \emph{\bibinfo{journal}{Nat. Photon.}}
    \textbf{\bibinfo{volume}{8}}, \bibinfo{pages}{821-829}
    (\bibinfo{year}{2014}).
    \bibitem{Ozawa2019}
    \bibinfo{author}{Ozawa, T.} \emph{et~al.}
    \newblock \bibinfo{title}{Topological photonics}.
    \newblock \emph{\bibinfo{journal}{Rev. Mod. Phys.}}
    \textbf{\bibinfo{volume}{91}}, \bibinfo{pages}{015006}
    (\bibinfo{year}{2019}).
    \bibitem{Vicencio2024}
    \bibinfo{author}{C\'{a}ceres-Aravena, G.} \emph{et~al.}
    \newblock \bibinfo{title}{Compact topological edge states in flux-dressed graphenelike photonic lattices}.
    \newblock \emph{\bibinfo{journal}{Phys. Rev. Lett.}}
    \textbf{\bibinfo{volume}{133}}, \bibinfo{pages}{116304}
    (\bibinfo{year}{2024}).
    \bibitem{Qu2024}
    \bibinfo{author}{Qu, T.} \emph{et~al.}
    \newblock \bibinfo{title}{Topological photonic alloy}.
    \newblock \emph{\bibinfo{journal}{Phys. Rev. Lett.}}
    \textbf{\bibinfo{volume}{132}}, \bibinfo{pages}{223802}
    (\bibinfo{year}{2024}).
    \bibitem{Rechtsman2024}
    \bibinfo{author}{Szameit, A.} \& 
    \bibinfo{author}{Rechtsman, M. C.}
    \newblock \bibinfo{title}{Discrete nonlinear topological photonics}.
    \newblock \emph{\bibinfo{journal}{Nat. Phys.}}
    \textbf{\bibinfo{volume}{20}}, \bibinfo{pages}{905-912}
    (\bibinfo{year}{2024}).
    \bibitem{Huber2016}
    \bibinfo{author}{Huber, S. D.}
    \newblock \bibinfo{title}{Topological mechanics}.
    \newblock \emph{\bibinfo{journal}{Nat. Phys.}}
    \textbf{\bibinfo{volume}{12}}, \bibinfo{pages}{621}
    (\bibinfo{year}{2016}).
    \bibitem{Raghu2008}
    \bibinfo{author}{Haldane, F. D. M.} \& 
    \bibinfo{author}{Raghu, S.}
    \newblock \bibinfo{title}{Possible realization of directional optical waveguides in photonic crystals with broken time-reversal symmetry}.
    \newblock \emph{\bibinfo{journal}{Phys. Rev. Lett.}}
    \textbf{\bibinfo{volume}{100}}, \bibinfo{pages}{013904}
    (\bibinfo{year}{2008}).
    \bibitem{Wangz2008}
    \bibinfo{author}{Wang, Z.},
    \bibinfo{author}{Chong, Y. D.},
    \bibinfo{author}{Joannopoulos, J. D.} \& 
    \bibinfo{author}{Solja\v{c}i\'{c}, M.}
    \newblock \bibinfo{title}{Reflection-free one-way edge modes in a gyromagnetic photonic crystal}.
    \newblock \emph{\bibinfo{journal}{Phys. Rev. Lett.}}
    \textbf{\bibinfo{volume}{100}}, \bibinfo{pages}{013905}
    (\bibinfo{year}{2008}).
    \bibitem{Wangz2009}
    \bibinfo{author}{Wang, Z.},
    \bibinfo{author}{Chong, Y.}, 
    \bibinfo{author}{Joannopoulos, J. D.} \& 
    \bibinfo{author}{Solja\v{c}i\'{c}, M.}
    \newblock \bibinfo{title}{Observation of unidirectional backscattering-immune topological electromagnetic states}.
    \newblock \emph{\bibinfo{journal}{Nature}}
    \textbf{\bibinfo{volume}{461}}, \bibinfo{pages}{772-775}
    (\bibinfo{year}{2009}).
    \bibitem{Chong2017}
    \bibinfo{author}{Mansha, S.} \& 
    \bibinfo{author}{Chong, Y. D.}
    \newblock \bibinfo{title}{Robust edge states in amorphous gyromagnetic photonic lattices}.
    \newblock \emph{\bibinfo{journal}{Phys. Rev. B}}
    \textbf{\bibinfo{volume}{96}}, \bibinfo{pages}{121405(R)}
    (\bibinfo{year}{2017}).
    \bibitem{Zhangs2017}
    \bibinfo{author}{Liu, C.},
    \bibinfo{author}{Gao, W.}, 
    \bibinfo{author}{Yang, B.} \& 
    \bibinfo{author}{Zhang, S.}
    \newblock \bibinfo{title}{Disorder-induced topological state transition in photonic metamaterials}.
    \newblock \emph{\bibinfo{journal}{Phys. Rev. Lett.}}
    \textbf{\bibinfo{volume}{119}}, \bibinfo{pages}{183901}
    (\bibinfo{year}{2017}).
    \bibitem{Shen2009}
    \bibinfo{author}{Li, J.},
    \bibinfo{author}{Chu, R.-L.}, 
    \bibinfo{author}{Jain, J. K.} \& 
    \bibinfo{author}{Shen, S.-Q.}
    \newblock \bibinfo{title}{Topological anderson insulator}.
    \newblock \emph{\bibinfo{journal}{Phys. Rev. Lett.}}
    \textbf{\bibinfo{volume}{102}}, \bibinfo{pages}{136806}
    (\bibinfo{year}{2009}).
    \bibitem{Groth2009}
    \bibinfo{author}{Groth, C. W.},
    \bibinfo{author}{Wimmer, M.}, 
    \bibinfo{author}{Akhmerov, A. R.},
    \bibinfo{author}{Tworzyd\l{}o, J.} \& 
    \bibinfo{author}{Beenakker, C. W. J.}
    \newblock \bibinfo{title}{Theory of the topological Anderson insulator}.
    \newblock \emph{\bibinfo{journal}{Phys. Rev. Lett.}}
    \textbf{\bibinfo{volume}{103}}, \bibinfo{pages}{196805}
    (\bibinfo{year}{2009}).
    \bibitem{Wang2011}
    \bibinfo{author}{Xing, Y.},
    \bibinfo{author}{Zhang, L.} \& 
    \bibinfo{author}{Wang, J.}
    \newblock \bibinfo{title}{Topological Anderson insulator phenomena}.
    \newblock \emph{\bibinfo{journal}{Phys. Rev. B}}
    \textbf{\bibinfo{volume}{84}}, \bibinfo{pages}{035110}
    (\bibinfo{year}{2011}).
    \bibitem{Shen2012}
    \bibinfo{author}{Zhang, Y.-Y.},
    \bibinfo{author}{Chu, R.-L.}, 
    \bibinfo{author}{Zhang, F.-C.} \& 
    \bibinfo{author}{Shen, S.-Q.}
    \newblock \bibinfo{title}{Localization and mobility gap in the topological Anderson insulator}.
    \newblock \emph{\bibinfo{journal}{Phys. Rev. B}}
    \textbf{\bibinfo{volume}{85}}, \bibinfo{pages}{035107}
    (\bibinfo{year}{2012}).
    \bibitem{Titum2015}
    \bibinfo{author}{Titum, P.},
    \bibinfo{author}{Lindner, N. H.}, 
    \bibinfo{author}{Rechtsman, M. C.} \& 
    \bibinfo{author}{Refael, G.}
    \newblock \bibinfo{title}{Disorder-induced floquet topological insulators}.
    \newblock \emph{\bibinfo{journal}{Phys. Rev. Lett.}}
    \textbf{\bibinfo{volume}{114}}, \bibinfo{pages}{056801}
    (\bibinfo{year}{2015}).
    \bibitem{Orth2016}
    \bibinfo{author}{Orth, C. P.},
    \bibinfo{author}{Sekera, T.}, 
    \bibinfo{author}{Bruder, C.} \& 
    \bibinfo{author}{Schmidt, T. L.}
    \newblock \bibinfo{title}{The topological Anderson insulator phase in the Kane-Mele model}.
    \newblock \emph{\bibinfo{journal}{Sci. Rep.}}
    \textbf{\bibinfo{volume}{6}}, \bibinfo{pages}{24007}
    (\bibinfo{year}{2016}).
    \bibitem{Jiang2019}
    \bibinfo{author}{Zhang, Z.-Q.},
    \bibinfo{author}{Wu, B.-L.}, 
    \bibinfo{author}{Song, J.} \& 
    \bibinfo{author}{Jiang, H.}
    \newblock \bibinfo{title}{Topological Anderson insulator in electric circuits}.
    \newblock \emph{\bibinfo{journal}{Phys. Rev. B}}
    \textbf{\bibinfo{volume}{100}}, \bibinfo{pages}{184202}
    (\bibinfo{year}{2019}).
    \bibitem{Huang2022}
    \bibinfo{author}{Wang, C.},
    \bibinfo{author}{Cheng, T.}, 
    \bibinfo{author}{Liu, Z.},
    \bibinfo{author}{Liu, F.} \& 
    \bibinfo{author}{Huang, H.}
    \newblock \bibinfo{title}{Structural amorphization-induced topological order}.
    \newblock \emph{\bibinfo{journal}{Phys. Rev. Lett.}}
    \textbf{\bibinfo{volume}{128}}, \bibinfo{pages}{056401}
    (\bibinfo{year}{2022}).
    \bibitem{Cui2022}
    \bibinfo{author}{Cui, X.},
    \bibinfo{author}{Zhang, R.-Y.}, 
    \bibinfo{author}{Zhang, Z.-Q.} \& 
    \bibinfo{author}{Chan, C. T.}
    \newblock \bibinfo{title}{Photonic ${\mathbb{Z}}_{2}$ topological Anderson insulators}.
    \newblock \emph{\bibinfo{journal}{Phys. Rev. Lett.}}
    \textbf{\bibinfo{volume}{129}}, \bibinfo{pages}{043902}
    (\bibinfo{year}{2022}).
    \bibitem{Sun2024}
    \bibinfo{author}{Ren, M.} \emph{et~al.}
    \newblock \bibinfo{title}{Realization of gapped and ungapped photonic topological Anderson insulators}.
    \newblock \emph{\bibinfo{journal}{Phys. Rev. Lett.}}
    \textbf{\bibinfo{volume}{132}}, \bibinfo{pages}{066602}
    (\bibinfo{year}{2024}).
    \bibitem{Luzp2025}
    \bibinfo{author}{Lu, Z.},
    \bibinfo{author}{Zhang, Y.} \& 
    \bibinfo{author}{Xu, Z.}
    \newblock \bibinfo{title}{Reentrant localization transitions in a topological Anderson insulator: A study of a generalized Su-Schrieffer-Heeger quasicrystal}.
    \newblock \emph{\bibinfo{journal}{Front. Phys.}}
    \textbf{\bibinfo{volume}{20}}, \bibinfo{pages}{024204}
    (\bibinfo{year}{2025}).
    \bibitem{Xie2009}
    \bibinfo{author}{Jiang, H.},
    \bibinfo{author}{Wang, L.}, 
    \bibinfo{author}{Sun, Q.-F.} \& 
    \bibinfo{author}{Xie, X. C.}
    \newblock \bibinfo{title}{Numerical study of the topological Anderson insulator in HgTe/CdTe quantum wells}.
    \newblock \emph{\bibinfo{journal}{Phys. Rev. B}}
    \textbf{\bibinfo{volume}{80}}, \bibinfo{pages}{165316}
    (\bibinfo{year}{2009}).
    \bibitem{Schmiedt2014}
    \bibinfo{author}{Altland, A.},
    \bibinfo{author}{Bagrets, D.}, 
    \bibinfo{author}{Fritz, L.},
    \bibinfo{author}{Kamenev, A.} \& 
    \bibinfo{author}{Schmiedt, H.}
    \newblock \bibinfo{title}{Quantum criticality of quasi-one-dimensional topological Anderson insulators}.
    \newblock \emph{\bibinfo{journal}{Phys. Rev. Lett.}}
    \textbf{\bibinfo{volume}{112}}, \bibinfo{pages}{206602}
    (\bibinfo{year}{2014}).
    \bibitem{Gadway2018}
    \bibinfo{author}{Meier, E. J.} \emph{et~al.}
    \newblock \bibinfo{title}{Observation of the topological Anderson insulator in disordered atomic wires}.
    \newblock \emph{\bibinfo{journal}{Science}}
    \textbf{\bibinfo{volume}{362}}, \bibinfo{pages}{929-933}
    (\bibinfo{year}{2018}).
    \bibitem{Szameit2018}
    \bibinfo{author}{St\"{u}tzer, S.} \emph{et~al.}
    \newblock \bibinfo{title}{Photonic topological Anderson insulators}.
    \newblock \emph{\bibinfo{journal}{Nature}}
    \textbf{\bibinfo{volume}{560}}, \bibinfo{pages}{461-465}
    (\bibinfo{year}{2018}).
    \bibitem{LiuGG2020}
    \bibinfo{author}{Liu, G.-G.} \emph{et~al.}
    \newblock \bibinfo{title}{Topological Anderson insulator in disordered photonic crystals}.
    \newblock \emph{\bibinfo{journal}{Phys. Rev. Lett.}}
    \textbf{\bibinfo{volume}{125}}, \bibinfo{pages}{133603}
    (\bibinfo{year}{2020}).
    \bibitem{XueP2022}
    \bibinfo{author}{Lin, Q.} \emph{et~al.}
    \newblock \bibinfo{title}{Observation of non-Hermitian topological Anderson insulator in quantum dynamics}.
    \newblock \emph{\bibinfo{journal}{Nat. Commun.}}
    \textbf{\bibinfo{volume}{13}}, \bibinfo{pages}{3229}
    (\bibinfo{year}{2022}).
    \bibitem{Dunlap1990}
    \bibinfo{author}{Dunlap, D. H.},
    \bibinfo{author}{Wu, H.-L.} \& 
    \bibinfo{author}{Phillips, P. W.}
    \newblock \bibinfo{title}{Absence of localization in a random-dimer model}.
    \newblock \emph{\bibinfo{journal}{Phys. Rev. Lett.}}
    \textbf{\bibinfo{volume}{65}}, \bibinfo{pages}{88}
    (\bibinfo{year}{1990}).
    \bibitem{ZhangDW202202}
    \bibinfo{author}{Liu, S.-N.},
    \bibinfo{author}{Zhang, G.-Q.},
    \bibinfo{author}{Tang, L.-Z.} \& 
    \bibinfo{author}{Zhang, D.-W.}
    \newblock \bibinfo{title}{Topological Anderson insulators induced by random binary disorders}.
    \newblock \emph{\bibinfo{journal}{Phys. Lett. A}}
    \textbf{\bibinfo{volume}{431}}, \bibinfo{pages}{128004}
    (\bibinfo{year}{2022}).
    \bibitem{Nava2023}
    \bibinfo{author}{Nava, A.},
    \bibinfo{author}{Campagnano, G.},
    \bibinfo{author}{Sodano, P.} \& 
    \bibinfo{author}{Giuliano, D.}
    \newblock \bibinfo{title}{Lindblad master equation approach to the topological phase transition in the disordered Su-Schrieffer-Heeger model}.
    \newblock \emph{\bibinfo{journal}{Phys. Rev. B}}
    \textbf{\bibinfo{volume}{107}}, \bibinfo{pages}{035113}
    (\bibinfo{year}{2023}).
    \bibitem{Cinnirella2024}
    \bibinfo{author}{Cinnirella, E. G.}, 
    \bibinfo{author}{Nava, A.},
    \bibinfo{author}{Campagnano, G.} \& 
    \bibinfo{author}{Giuliano, D.}
    \newblock \bibinfo{title}{Fate of high winding number topological phases in the disordered extended Su-Schrieffer-Heeger model}.
    \newblock \emph{\bibinfo{journal}{Phys. Rev. B}}
    \textbf{\bibinfo{volume}{109}}, \bibinfo{pages}{035114}
    (\bibinfo{year}{2024}).
    \bibitem{KangDW2024}
    \bibinfo{author}{Zuo, Z.-W.},
    \bibinfo{author}{Lin, J.-R.} \& 
    \bibinfo{author}{Kang, D.}
    \newblock \bibinfo{title}{Topological inverse Anderson insulator}.
    \newblock \emph{\bibinfo{journal}{Phys. Rev. B}}
    \textbf{\bibinfo{volume}{110}}, \bibinfo{pages}{085157}
    (\bibinfo{year}{2024}).
    \bibitem{GhoshAK2024}
    \bibinfo{author}{Ghosh, A. K.},
    \bibinfo{author}{Nag, T.} \& 
    \bibinfo{author}{Saha, A.}
    \newblock \bibinfo{title}{Floquet second-order topological Anderson insulator hosting corner localized modes}.
    \newblock \emph{\bibinfo{journal}{Phys. Rev. B}}
    \textbf{\bibinfo{volume}{110}}, \bibinfo{pages}{125427}
    (\bibinfo{year}{2024}).
    \bibitem{ZhuSL2021}
    \bibinfo{author}{Zhang, G.-Q.},
    \bibinfo{author}{Tang, L.-Z.}, 
    \bibinfo{author}{Zhang, L.-F.},
    \bibinfo{author}{Zhang, D.-W.} \& 
    \bibinfo{author}{Zhu, S.-L.}
    \newblock \bibinfo{title}{Connecting topological Anderson and Mott insulators in disordered interacting fermionic systems}.
    \newblock \emph{\bibinfo{journal}{Phys. Rev. B}}
    \textbf{\bibinfo{volume}{104}}, \bibinfo{pages}{L161118}
    (\bibinfo{year}{2021}).
    \bibitem{ZhangDW2022}
    \bibinfo{author}{Tang, L.-Z.},
    \bibinfo{author}{Liu, S.-N.}, 
    \bibinfo{author}{Zhang, G.-Q.} \& 
    \bibinfo{author}{Zhang, D.-W.}
    \newblock \bibinfo{title}{Topological Anderson insulators with different bulk states in quasiperiodic chains}.
    \newblock \emph{\bibinfo{journal}{Phys. Rev. A}}
    \textbf{\bibinfo{volume}{105}}, \bibinfo{pages}{063327}
    (\bibinfo{year}{2022}).
    \bibitem{LuZP2022}
    \bibinfo{author}{Lu, Z.},
    \bibinfo{author}{Xu, Z.} \& 
    \bibinfo{author}{Zhang, Y.}
    \newblock \bibinfo{title}{Exact mobility edges and topological Anderson insulating phase in a slowly varying quasiperiodic model}.
    \newblock \emph{\bibinfo{journal}{Ann. Phys.}}
    \textbf{\bibinfo{volume}{534}}, \bibinfo{pages}{2200203}
    (\bibinfo{year}{2022}).
    \bibitem{Padhan2024}
    \bibinfo{author}{Padhan, A.},
    \bibinfo{author}{Padhi, S. R.} \& 
    \bibinfo{author}{Mishra, T.}
    \newblock \bibinfo{title}{Complete delocalization and reentrant topological transition in a non-Hermitian quasiperiodic lattice}.
    \newblock \emph{\bibinfo{journal}{Phys. Rev. B}}
    \textbf{\bibinfo{volume}{109}}, \bibinfo{pages}{L020203}
    (\bibinfo{year}{2024}).
    \bibitem{XueQK2024}
    \bibinfo{author}{Li, X.} \emph{et~al.}
    \newblock \bibinfo{title}{Mapping the topology-localization phase diagram with quasiperiodic disorder using a programmable superconducting simulator}.
    \newblock \emph{\bibinfo{journal}{Phys. Rev. Res.}}
    \textbf{\bibinfo{volume}{6}}, \bibinfo{pages}{L042038}
    (\bibinfo{year}{2024}).
	\bibitem{Ouyang2024}
	\bibinfo{author}{Ouyang, C.},
	\bibinfo{author}{He, Q.}, 
	\bibinfo{author}{Xu, D.-H.} \& 
	\bibinfo{author}{Liu, F.} 
	\newblock \bibinfo{title}{Higher-order topology in Fibonacci quasicrystals}. 
	\newblock \emph{\bibinfo{journal}{Phys. Rev. B}} \textbf{\bibinfo{volume}{110}}, \bibinfo{pages}{075425} (\bibinfo{year}{2024}).
    \bibitem{Garreau2008}
    \bibinfo{author}{Chab\'{e}, J.} \emph{et~al.}
    \newblock \bibinfo{title}{Experimental Observation of the Anderson metal-insulator transition with atomic matter waves}.
    \newblock \emph{\bibinfo{journal}{Phys. Rev. Lett.}}
    \textbf{\bibinfo{volume}{101}}, \bibinfo{pages}{255702}
    (\bibinfo{year}{2008}).
    \bibitem{Modugno2014}
    \bibinfo{author}{D'Errico, C.} \emph{et~al.}
    \newblock \bibinfo{title}{Observation of a disordered bosonic insulator from weak to strong interactions}.
    \newblock \emph{\bibinfo{journal}{Phys. Rev. Lett.}}
    \textbf{\bibinfo{volume}{113}}, \bibinfo{pages}{095301}
    (\bibinfo{year}{2014}).
    \bibitem{YaoH2020}
    \bibinfo{author}{Yao, H.},
    \bibinfo{author}{Giamarchi, T.} \& 
    \bibinfo{author}{Sanchez-Palencia, L.}
    \newblock \bibinfo{title}{Lieb-Liniger bosons in a shallow quasiperiodic potential: Bose glass phase and fractal Mott lobes}.
    \newblock \emph{\bibinfo{journal}{Phys. Rev. Lett.}}
    \textbf{\bibinfo{volume}{125}}, \bibinfo{pages}{060401}
    (\bibinfo{year}{2020}).
    \bibitem{DaiQ2023}
    \bibinfo{author}{Dai, Q.},
    \bibinfo{author}{Lu, Z.} \& 
    \bibinfo{author}{Xu, Z.}
    \newblock \bibinfo{title}{Emergence of multifractality through cascadelike transitions in a mosaic interpolating Aubry-Andr\'e-Fibonacci chain}.
    \newblock \emph{\bibinfo{journal}{Phys. Rev. B}}
    \textbf{\bibinfo{volume}{108}}, \bibinfo{pages}{144207}
    (\bibinfo{year}{2023}).
    \bibitem{Rai2021}
    \bibinfo{author}{Rai, G.},
    \bibinfo{author}{Schl\"{o}mer, H.}, 
    \bibinfo{author}{Matsumura, C.},
    \bibinfo{author}{Haas, S.} \& 
    \bibinfo{author}{Jagannathan, A.}
    \newblock \bibinfo{title}{Bulk topological signatures of a quasicrystal}.
    \newblock \emph{\bibinfo{journal}{Phys. Rev. B}}
    \textbf{\bibinfo{volume}{104}}, \bibinfo{pages}{184202}
    (\bibinfo{year}{2021}).
    \bibitem{Haas2024}
    \bibinfo{author}{Wang, Y.},
    \bibinfo{author}{Rai, G.}, 
    \bibinfo{author}{Matsumura, C.},
    \bibinfo{author}{Jagannathan, A.} \& 
    \bibinfo{author}{Haas, S.}
    \newblock \bibinfo{title}{Superconductivity in the Fibonacci chain}.
    \newblock \emph{\bibinfo{journal}{Phys. Rev. B}}
    \textbf{\bibinfo{volume}{109}}, \bibinfo{pages}{214507}
    (\bibinfo{year}{2024}).
    \bibitem{Kobialka2024}
    \bibinfo{author}{Kobia\l{}ka, A.} \emph{et~al.}
    \newblock \bibinfo{title}{Topological superconductivity in Fibonacci quasicrystals}.
    \newblock \emph{\bibinfo{journal}{Phys. Rev. B}}
    \textbf{\bibinfo{volume}{110}}, \bibinfo{pages}{134508}
    (\bibinfo{year}{2024}).
    \bibitem{Sandberg2024}
    \bibinfo{author}{Sandberg, A.},
    \bibinfo{author}{Awoga, O. A.}, 
    \bibinfo{author}{Black-Schaffer, A. M.} \& 
    \bibinfo{author}{Holmvall, P.}
    \newblock \bibinfo{title}{Josephson effect in a Fibonacci quasicrystal}.
    \newblock \emph{\bibinfo{journal}{Phys. Rev. B}}
    \textbf{\bibinfo{volume}{110}}, \bibinfo{pages}{104513}
    (\bibinfo{year}{2024}).
    \bibitem{Sbroscia2020}
    \bibinfo{author}{Sbroscia, M.} \emph{et~al.}
    \newblock \bibinfo{title}{Observing localization in a 2D quasicrystalline optical lattice}.
    \newblock \emph{\bibinfo{journal}{Phys. Rev. Lett.}}
    \textbf{\bibinfo{volume}{125}}, \bibinfo{pages}{200604}
    (\bibinfo{year}{2020}).
    \bibitem{Jagannathan2021}
    \bibinfo{author}{Jagannathan, A.}
    \newblock \bibinfo{title}{The Fibonacci quasicrystal: Case study of hidden dimensions and multifractality}.
    \newblock \emph{\bibinfo{journal}{Rev. Mod. Phys.}}
    \textbf{\bibinfo{volume}{93}}, \bibinfo{pages}{045001}
    (\bibinfo{year}{2021}).
   \bibitem{Liutong2025}
   \bibinfo{author}{Liu, T.}
   \newblock \bibinfo{title}{Dual-space invariance as a definitive signature of critical states in Anderson localization.} \newblock \eprint{arXiv:2411.09067v4}
   (\bibinfo{year}{2025}).
    \bibitem{1983}
    \bibinfo{author}{Kohmoto, M.},
    \bibinfo{author}{Kadanoff, L. P.} \& 
    \bibinfo{author}{Tang, C.}
    \newblock \bibinfo{title}{Localization problem in one dimension: mapping and escape}.
    \newblock \emph{\bibinfo{journal}{Phys. Rev. Lett.}}
    \textbf{\bibinfo{volume}{50}}, \bibinfo{pages}{1870}
    (\bibinfo{year}{1983}).
    \bibitem{Zilberberg2012}
    \bibinfo{author}{Kraus, Y. E.} \& 
    \bibinfo{author}{Zilberberg, O.}
    \newblock \bibinfo{title}{Topological equivalence between the Fibonacci quasicrystal and the Harper model}.
    \newblock \emph{\bibinfo{journal}{Phys. Rev. Lett.}}
    \textbf{\bibinfo{volume}{109}}, \bibinfo{pages}{116404}
    (\bibinfo{year}{2012}).
    \bibitem{Verbin2013}
    \bibinfo{author}{Verbin, M.},
    \bibinfo{author}{Zilberberg, O.}, 
    \bibinfo{author}{Kraus, Y. E.},
    \bibinfo{author}{Lahini, Y.} \& 
    \bibinfo{author}{Silberberg, Y.}
    \newblock \bibinfo{title}{Observation of topological phase transitions in photonic quasicrystals}.
    \newblock \emph{\bibinfo{journal}{Phys. Rev. Lett.}}
    \textbf{\bibinfo{volume}{110}}, \bibinfo{pages}{076403}
    (\bibinfo{year}{2013}).
    \bibitem{Mace2016}
    \bibinfo{author}{Mac\'{e}, N.},
    \bibinfo{author}{Jagannathan, A.} \& 
    \bibinfo{author}{Pi\'{e}chon, F.}
    \newblock \bibinfo{title}{Fractal dimensions of wave functions and local spectral measures on the Fibonacci chain}.
    \newblock \emph{\bibinfo{journal}{Phys. Rev. B}}
    \textbf{\bibinfo{volume}{93}}, \bibinfo{pages}{205153}
    (\bibinfo{year}{2016}).
    \bibitem{Mace2017}
    \bibinfo{author}{Mac\'{e}, N.},
    \bibinfo{author}{Jagannathan, A.}, 
    \bibinfo{author}{Kalugin, P.},
    \bibinfo{author}{Mosseri, R.} \& 
    \bibinfo{author}{Pi\'{e}chon, F.}
    \newblock \bibinfo{title}{Critical eigenstates and their properties in one- and two-dimensional quasicrystals}.
    \newblock \emph{\bibinfo{journal}{Phys. Rev. B}}
    \textbf{\bibinfo{volume}{96}}, \bibinfo{pages}{045138}
    (\bibinfo{year}{2017}).
    \bibitem{Bian2021}
    \bibinfo{author}{Liu, J.-Q.} \& 
    \bibinfo{author}{Bian, X.-B.}
    \newblock \bibinfo{title}{Multichannel high-order harmonic generation from fractal bands in Fibonacci quasicrystals}.
    \newblock \emph{\bibinfo{journal}{Phys. Rev. Lett.}}
    \textbf{\bibinfo{volume}{127}}, \bibinfo{pages}{213901}
    (\bibinfo{year}{2021}).
    \bibitem{Singh2015}
    \bibinfo{author}{Singh, K.},
    \bibinfo{author}{Saha, K.}, 
    \bibinfo{author}{Parameswaran, S. A.} \& 
    \bibinfo{author}{Weld, D. M.}
    \newblock \bibinfo{title}{Fibonacci optical lattices for tunable quantum quasicrystals}.
    \newblock \emph{\bibinfo{journal}{Phys. Rev. A}}
    \textbf{\bibinfo{volume}{92}}, \bibinfo{pages}{063426}
    (\bibinfo{year}{2015}).
    \bibitem{Weld2024}
    \bibinfo{author}{Shimasaki, T.} \emph{et~al.}
    \newblock \bibinfo{title}{Reversible phasonic control of a quantum phase transition in a quasicrystal}.
    \newblock \emph{\bibinfo{journal}{Phys. Rev. Lett.}}
    \textbf{\bibinfo{volume}{133}}, \bibinfo{pages}{083405}
    (\bibinfo{year}{2024}).
    \bibitem{Negro2003}
    \bibinfo{author}{Dal Negro, L.} \emph{et~al.}
    \newblock \bibinfo{title}{Light transport through the band-edge states of Fibonacci quasicrystals}.
    \newblock \emph{\bibinfo{journal}{Phys. Rev. Lett.}}
    \textbf{\bibinfo{volume}{90}}, \bibinfo{pages}{055501}
    (\bibinfo{year}{2003}).
    \bibitem{Kraus2012}
    \bibinfo{author}{Kraus, Y. E.},
    \bibinfo{author}{Lahini, Y.}, 
    \bibinfo{author}{Ringel, Z.},
    \bibinfo{author}{Verbin, M.} \& 
    \bibinfo{author}{Zilberberg, O.}
    \newblock \bibinfo{title}{Topological states and adiabatic pumping in quasicrystals}.
    \newblock \emph{\bibinfo{journal}{Phys. Rev. Lett.}}
    \textbf{\bibinfo{volume}{109}}, \bibinfo{pages}{106402}
    (\bibinfo{year}{2012}).
    \bibitem{Verbin2015}
    \bibinfo{author}{Verbin, M.},
    \bibinfo{author}{Zilberberg, O.}, 
    \bibinfo{author}{Lahini, Y.},
    \bibinfo{author}{Kraus, Y. E.} \& 
    \bibinfo{author}{Silberberg, Y.}
    \newblock \bibinfo{title}{Topological pumping over a photonic Fibonacci quasicrystal}.
    \newblock \emph{\bibinfo{journal}{Phys. Rev. B}}
    \textbf{\bibinfo{volume}{91}}, \bibinfo{pages}{064201}
    (\bibinfo{year}{2015}).
    \bibitem{Tanese2014}
    \bibinfo{author}{Tanese, D.} \emph{et~al.}
    \newblock \bibinfo{title}{Fractal energy spectrum of a polariton gas in a Fibonacci quasiperiodic potential}.
    \newblock \emph{\bibinfo{journal}{Phys. Rev. Lett.}}
    \textbf{\bibinfo{volume}{112}}, \bibinfo{pages}{146404}
    (\bibinfo{year}{2014}).
    \bibitem{Goblot2020}
    \bibinfo{author}{Goblot, V.} \emph{et~al.}
    \newblock \bibinfo{title}{Emergence of criticality through a cascade of delocalization transitions in quasiperiodic chains.}.
    \newblock \emph{\bibinfo{journal}{Nat. Phys.}}
    \textbf{\bibinfo{volume}{16}}, \bibinfo{pages}{832-836}
    (\bibinfo{year}{2020}).
    \bibitem{Reisner2023}
    \bibinfo{author}{Reisner, M.},
    \bibinfo{author}{Tahmi, Y.}, 
    \bibinfo{author}{Pi\'{e}chon, F.},
    \bibinfo{author}{Kuhl, U.} \& 
    \bibinfo{author}{Mortessagne, F.}
    \newblock \bibinfo{title}{Experimental observation of multifractality in Fibonacci chains}.
    \newblock \emph{\bibinfo{journal}{Phys. Rev. B}}
    \textbf{\bibinfo{volume}{108}}, \bibinfo{pages}{064210}
    (\bibinfo{year}{2023}).
    \bibitem{Franca2024}
    \bibinfo{author}{Franca, S.},
    \bibinfo{author}{Seidemann, T.}, 
    \bibinfo{author}{Hassler, F.},
    \bibinfo{author}{van den Brink, J.} \& 
    \bibinfo{author}{Fulga, I. C.}
    \newblock \bibinfo{title}{Impedance spectroscopy of chiral symmetric topoelectrical circuits}.
    \newblock \emph{\bibinfo{journal}{Phys. Rev. B}}
    \textbf{\bibinfo{volume}{109}}, \bibinfo{pages}{L241103}
    (\bibinfo{year}{2024}).
    \bibitem{Dareau2017}
    \bibinfo{author}{Dareau, A.} \emph{et~al.}
    \newblock \bibinfo{title}{Revealing the topology of quasicrystals with a diffraction experiment}.
    \newblock \emph{\bibinfo{journal}{Phys. Rev. Lett.}}
    \textbf{\bibinfo{volume}{119}}, \bibinfo{pages}{215304}
    (\bibinfo{year}{2017}).
    \bibitem{LiuF2018}
    \bibinfo{author}{Huang, H.} \& 
    \bibinfo{author}{Liu, F.}
    \newblock \bibinfo{title}{Quantum spin Hall effect and spin Bott index in a quasicrystal lattice}.
    \newblock \emph{\bibinfo{journal}{Phys. Rev. Lett.}}
    \textbf{\bibinfo{volume}{121}}, \bibinfo{pages}{126401}
    (\bibinfo{year}{2018}).
    \bibitem{LiuF2019}
    \bibinfo{author}{Huang, H.} \& 
    \bibinfo{author}{Liu, F.}
    \newblock \bibinfo{title}{Comparison of quantum spin Hall states in quasicrystals and crystals}.
    \newblock \emph{\bibinfo{journal}{Phys. Rev. B}}
    \textbf{\bibinfo{volume}{100}}, \bibinfo{pages}{085119}
    (\bibinfo{year}{2019}).
    \bibitem{XuDH2020}
    \bibinfo{author}{Chen, R.},
    \bibinfo{author}{Chen, C.-Z.}, 
    \bibinfo{author}{Gao, J.-H.},
    \bibinfo{author}{Zhou, B.} \& 
    \bibinfo{author}{Xu, D.-H.}
    \newblock \bibinfo{title}{Higher-order topological insulators in quasicrystals}.
    \newblock \emph{\bibinfo{journal}{Phys. Rev. Lett.}}
    \textbf{\bibinfo{volume}{124}}, \bibinfo{pages}{036803}
    (\bibinfo{year}{2020}).
    \bibitem{Kohmoto1983}
    \bibinfo{author}{Kohmoto, M.}
    \newblock \bibinfo{title}{Metal-insulator transition and scaling for incommensurate systems}.
    \newblock \emph{\bibinfo{journal}{Phys. Rev. Lett.}}
    \textbf{\bibinfo{volume}{51}}, \bibinfo{pages}{1198}
    (\bibinfo{year}{1983}).
    \bibitem{LiuXJ2013}
    \bibinfo{author}{Liu, X.-J.},
    \bibinfo{author}{Liu, Z.-X.} \& 
    \bibinfo{author}{Cheng, M.}
    \newblock \bibinfo{title}{Manipulating topological edge spins in a one-dimensional optical lattice}.
    \newblock \emph{\bibinfo{journal}{Phys. Rev. Lett.}}
    \textbf{\bibinfo{volume}{110}}, \bibinfo{pages}{076401}
    (\bibinfo{year}{2013}).
    \bibitem{Wang2020}
    \bibinfo{author}{Wang, Y.},
    \bibinfo{author}{Zhang, L.}, 
    \bibinfo{author}{Niu, S.},
    \bibinfo{author}{Yu, D.} \& 
    \bibinfo{author}{Liu, X.-J.}
    \newblock \bibinfo{title}{Realization and detection of nonergodic critical phases in an optical Raman lattice}.
    \newblock \emph{\bibinfo{journal}{Phys. Rev. Lett.}}
    \textbf{\bibinfo{volume}{125}}, \bibinfo{pages}{073204}
    (\bibinfo{year}{2020}).
    \bibitem{Gadway2015}
    \bibinfo{author}{Gadway, B.}
    \newblock \bibinfo{title}{Atom-optics approach to studying transport phenomena}.
    \newblock \emph{\bibinfo{journal}{Phys. Rev. A}}
    \textbf{\bibinfo{volume}{92}}, \bibinfo{pages}{043606}
    (\bibinfo{year}{2015}).
    \bibitem{Gadway2016}
    \bibinfo{author}{Meier, E. J.},
    \bibinfo{author}{An, F. A.} \& 
    \bibinfo{author}{Gadway, B.}
    \newblock \bibinfo{title}{Atom-optics simulator of lattice transport phenomena}.
    \newblock \emph{\bibinfo{journal}{Phys. Rev. A}}
    \textbf{\bibinfo{volume}{93}}, \bibinfo{pages}{051602(R)}
    (\bibinfo{year}{2016}).
    \bibitem{Gadway2017}
    \bibinfo{author}{An, F. A.},
    \bibinfo{author}{Meier, E. J.} \& 
    \bibinfo{author}{Gadway, B.}
    \newblock \bibinfo{title}{Diffusive and arrested transport of atoms under tailored disorder}.
    \newblock \emph{\bibinfo{journal}{Nat. Commun.}}
    \textbf{\bibinfo{volume}{8}}, \bibinfo{pages}{325}
    (\bibinfo{year}{2017}).
    \bibitem{Gadway201802}
    \bibinfo{author}{An, F. A.},
    \bibinfo{author}{Meier, E. J.}, 
    \bibinfo{author}{Ang'ong'a, J.} \& 
    \bibinfo{author}{Gadway, B.}
    \newblock \bibinfo{title}{Correlated dynamics in a synthetic lattice of momentum states}.
    \newblock \emph{\bibinfo{journal}{Phys. Rev. Lett.}}
    \textbf{\bibinfo{volume}{120}}, \bibinfo{pages}{040407}
    (\bibinfo{year}{2018}).
    \bibitem{WangYF2022}
    \bibinfo{author}{Wang, Y.} \emph{et~al.}
    \newblock \bibinfo{title}{Observation of interaction-induced mobility edge in an atomic Aubry-Andr\'e wire}.
    \newblock \emph{\bibinfo{journal}{Phys. Rev. Lett.}}
    \textbf{\bibinfo{volume}{129}}, \bibinfo{pages}{103401}
    (\bibinfo{year}{2022}).
    \bibitem{LiYQ2022}
    \bibinfo{author}{Li, Y.} \emph{et~al.}
    \newblock \bibinfo{title}{Atom-optically synthetic gauge fields for a noninteracting Bose gas}.
    \newblock \emph{\bibinfo{journal}{Light. Sci. Appl.}}
    \textbf{\bibinfo{volume}{11}}, \bibinfo{pages}{13}
    (\bibinfo{year}{2022}).
    \bibitem{YanBo2022}
    \bibinfo{author}{Liang, Q.} \emph{et~al.}
    \newblock \bibinfo{title}{Dynamic signatures of non-Hermitian skin effect and topology in ultracold atoms}.
    \newblock \emph{\bibinfo{journal}{Phys. Rev. Lett.}}
    \textbf{\bibinfo{volume}{129}}, \bibinfo{pages}{070401}
    (\bibinfo{year}{2022}).
    \bibitem{YanBo202202}
    \bibinfo{author}{Li, H.} \emph{et~al.}
    \newblock \bibinfo{title}{Aharonov-Bohm caging and inverse Anderson transition in ultracold atoms}.
    \newblock \emph{\bibinfo{journal}{Phys. Rev. Lett.}}
    \textbf{\bibinfo{volume}{129}}, \bibinfo{pages}{220403}
    (\bibinfo{year}{2022}).
    \bibitem{LiYQ2023}
    \bibinfo{author}{Li, Y.} \emph{et~al.}
    \newblock \bibinfo{title}{Observation of frustrated chiral dynamics in an interacting triangular flux ladder}.
    \newblock \emph{\bibinfo{journal}{Nat. Commun.}}
    \textbf{\bibinfo{volume}{14}}, \bibinfo{pages}{7560}
    (\bibinfo{year}{2023}).
    \bibitem{YanBo2024}
    \bibinfo{author}{Liang, Q.} \emph{et~al.}
    \newblock \bibinfo{title}{Chiral dynamics of ultracold atoms under a tunable SU(2) synthetic gauge field}.
    \newblock \emph{\bibinfo{journal}{Nat. Phys.}}
    \textbf{\bibinfo{volume}{20}}, \bibinfo{pages}{1738-1743}
    (\bibinfo{year}{2024}).
    \bibitem{Gadway2024}
    \bibinfo{author}{Paladugu, S. N. M.},
    \bibinfo{author}{Chen, T.}, 
    \bibinfo{author}{An, F. A.},
    \bibinfo{author}{Yan, B.} \& 
    \bibinfo{author}{Gatway, B.}
    \newblock \bibinfo{title}{Injection spectroscopy of momentum state lattices}.
    \newblock \emph{\bibinfo{journal}{Commun. Phys.}}
    \textbf{\bibinfo{volume}{7}}, \bibinfo{pages}{39}
    (\bibinfo{year}{2024}).
    \bibitem{Massignan2017}
    \bibinfo{author}{Cardano, F.} \emph{et~al.}
    \newblock \bibinfo{title}{Detection of Zak phases and topological invariants in a chiral quantum walk of twisted photons}.
    \newblock \emph{\bibinfo{journal}{Nat. Commun.}}
    \textbf{\bibinfo{volume}{8}}, \bibinfo{pages}{15516}
    (\bibinfo{year}{2017}).
    \bibitem{ZhouBZ2019}
    \bibinfo{author}{Zhou, B.},
    \bibinfo{author}{Yang, C.} \& 
    \bibinfo{author}{Chen, S.}
    \newblock \bibinfo{title}{Signature of a nonequilibrium quantum phase transition in the long-time average of the Loschmidt echo}.
    \newblock \emph{\bibinfo{journal}{Phys. Rev. B}}
    \textbf{\bibinfo{volume}{100}}, \bibinfo{pages}{184313}
    (\bibinfo{year}{2019}).
    \bibitem{ZhouBZ2021}
    \bibinfo{author}{Zhou, B.},
    \bibinfo{author}{Zeng, Y.} \& 
    \bibinfo{author}{Chen, S.}
    \newblock \bibinfo{title}{Exact zeros of the Loschmidt echo and quantum speed limit time for the dynamical quantum phase transition in finite-size systems}.
    \newblock \emph{\bibinfo{journal}{Phys. Rev. B}}
    \textbf{\bibinfo{volume}{104}}, \bibinfo{pages}{094311}
    (\bibinfo{year}{2021}).
    \bibitem{Langari2021}
    \bibinfo{author}{Sadrzadeh, M.},
    \bibinfo{author}{Jafari, R.} \& 
    \bibinfo{author}{Langari, A.}
    \newblock \bibinfo{title}{Dynamical topological quantum phase transitions at criticality}.
    \newblock \emph{\bibinfo{journal}{Phys. Rev. B}}
    \textbf{\bibinfo{volume}{103}}, \bibinfo{pages}{144305}
    (\bibinfo{year}{2021}).
    \bibitem{Nehra2024}
    \bibinfo{author}{Nehra, R.} \& 
    \bibinfo{author}{Roy, D.}
    \newblock \bibinfo{title}{Anomalous dynamical response of non-Hermitian topological phases}.
    \newblock \emph{\bibinfo{journal}{Phys. Rev. B}}
    \textbf{\bibinfo{volume}{109}}, \bibinfo{pages}{094311}
    (\bibinfo{year}{2024}).
    \bibitem{ChenB2024}
    \bibinfo{author}{Xiao, H.-X.} \emph{et~al.}
    \newblock \bibinfo{title}{Dynamical topological quantum phase transitions in high-order topological systems}.
    \newblock \emph{\bibinfo{journal}{Phys. Rev. B}}
    \textbf{\bibinfo{volume}{110}}, \bibinfo{pages}{064306}
    (\bibinfo{year}{2024}).
    \bibitem{HuZX2024}
    \bibinfo{author}{Jing, Y.},
    \bibinfo{author}{Dong, J.-J.}, 
    \bibinfo{author}{Zhang, Y.-Y.} \& 
    \bibinfo{author}{Hu, Z.-X.}
    \newblock \bibinfo{title}{Biorthogonal dynamical quantum phase transitions in non-Hermitian systems}.
    \newblock \emph{\bibinfo{journal}{Phys. Rev. Lett.}}
    \textbf{\bibinfo{volume}{132}}, \bibinfo{pages}{220402}
    (\bibinfo{year}{2024}).
    \bibitem{Heyl2013}
    \bibinfo{author}{Heyl, M.},
    \bibinfo{author}{Polkovnikov, A.} \& 
    \bibinfo{author}{Kehrein, S.}
    \newblock \bibinfo{title}{Dynamical quantum phase transitions in the transverse-field Ising model}.
    \newblock \emph{\bibinfo{journal}{Phys. Rev. Lett.}}
    \textbf{\bibinfo{volume}{110}}, \bibinfo{pages}{135704}
    (\bibinfo{year}{2013}).
    \bibitem{Heyl2014}
    \bibinfo{author}{Heyl, M.}
    \newblock \bibinfo{title}{Dynamical quantum phase transitions in systems with broken-symmetry phases}.
    \newblock \emph{\bibinfo{journal}{Phys. Rev. Lett.}}
    \textbf{\bibinfo{volume}{113}}, \bibinfo{pages}{205701}
    (\bibinfo{year}{2014}).
    \bibitem{Heyl2015} 
    \bibinfo{author}{Heyl, M.}
    \newblock \bibinfo{title}{Scaling and universality at dynamical quantum phase transitions}.
    \newblock \emph{\bibinfo{journal}{Phys. Rev. Lett.}}
    \textbf{\bibinfo{volume}{115}}, \bibinfo{pages}{140602}
    (\bibinfo{year}{2015}).
    \bibitem{Jafari2017}
    \bibinfo{author}{Jafari, R.} \& 
    \bibinfo{author}{Johannesson, H.}
    \newblock \bibinfo{title}{Loschmidt echo revivals: Critical and noncritical}.
    \newblock \emph{\bibinfo{journal}{Phys. Rev. Lett.}}
    \textbf{\bibinfo{volume}{118}}, \bibinfo{pages}{015701}
    (\bibinfo{year}{2017}).
    \bibitem{Heyl2018}
    \bibinfo{author}{Heyl, M.},
    \bibinfo{author}{Pollmann, F.} \& 
    \bibinfo{author}{D\'{o}ra, B.}
    \newblock \bibinfo{title}{Detecting equilibrium and dynamical quantum phase transitions in Ising chains via out-of-time-ordered correlators}.
    \newblock \emph{\bibinfo{journal}{Phys. Rev. Lett.}}
    \textbf{\bibinfo{volume}{121}}, \bibinfo{pages}{016801}
    (\bibinfo{year}{2018}).
    \bibitem{LiJun2020}
    \bibinfo{author}{N, X.} \emph{et~al.}
    \newblock \bibinfo{title}{Experimental observation of equilibrium and dynamical quantum phase transitions via out-of-time-ordered correlators}.
    \newblock \emph{\bibinfo{journal}{Phys. Rev. Lett.}}
    \textbf{\bibinfo{volume}{124}}, \bibinfo{pages}{250601}
    (\bibinfo{year}{2020}).
    \bibitem{Bandyopadhyay2021}
    \bibinfo{author}{Bandyopadhyay, S.},
    \bibinfo{author}{Polkovnikov, A.} \& 
    \bibinfo{author}{Dutta, A.}
    \newblock \bibinfo{title}{Observing dynamical quantum phase transitions through quasilocal string operators}.
    \newblock \emph{\bibinfo{journal}{Phys. Rev. Lett.}}
    \textbf{\bibinfo{volume}{126}}, \bibinfo{pages}{200602}
    (\bibinfo{year}{2021}).
    \bibitem{Rosch2012}
    \bibinfo{author}{Schneider, U.} \emph{et~al.}
    \newblock \bibinfo{title}{Fermionic transport and out-of-equilibrium dynamics in a homogeneous Hubbard model with ultracold atoms}.
    \newblock \emph{\bibinfo{journal}{Nat. Phys.}}
    \textbf{\bibinfo{volume}{8}}, \bibinfo{pages}{213-218}
    (\bibinfo{year}{2012}).
    \bibitem{BlochI2018}
    \bibinfo{author}{L\"{u}schen, H. P.} \emph{et~al.}
    \newblock \bibinfo{title}{Single-particle mobility edge in a one-dimensional quasiperiodic optical lattice}.
    \newblock \emph{\bibinfo{journal}{Phys. Rev. Lett.}}
    \textbf{\bibinfo{volume}{120}}, \bibinfo{pages}{160404}
    (\bibinfo{year}{2018}).
\bibitem{Zhang2016}
\bibinfo{author}{Zhang, P.} \& 
\bibinfo{author}{Nori, F.}
\newblock \bibinfo{title}{Majorana bound states in a disordered quantum dot chain}.
\newblock \emph{\bibinfo{journal}{New J. Phys.}}
\textbf{\bibinfo{volume}{18}}, \bibinfo{pages}{043033}
(\bibinfo{year}{2016}).

\end{thebibliography}

\begin{thebibliography}{1}
	\par
	\par
\bibitem{Zilberberg2012fulu}
\bibinfo{author}{Kraus, Y. E.} \& 
\bibinfo{author}{Zilberberg, O.}
\newblock \bibinfo{title}{Topological equivalence between the Fibonacci quasicrystal and the Harper model}.
\newblock \emph{\bibinfo{journal}{Phys. Rev. Lett.}}
\textbf{\bibinfo{volume}{109}}, \bibinfo{pages}{116404}
(\bibinfo{year}{2012}).
\end{thebibliography}

\section*{Data Availability}
The data that support the findings of this study are available from the corresponding authors upon reasonable request.

\section*{Code availability}
The code used for the analysis is available from the authors upon reasonable request.

\section*{Acknowledgements}
Z.X. is supported by the NSFC (Grant No. 12375016 and No. 12461160324), and Beijing National Laboratory for Condensed Matter Physics (No. 2023BNLCMPKF001). This work is also supported by NSF for Shanxi Province (Grant No. 1331KSC).

\section*{Author contributions statement}
Z.H.X. and R.J.J. conceived and designed the project. R.J.J. performed the numerical simulations.
Z.H.X. provided the explanation of the numerical results. All authors contributed to the discussion of the results and wrote the paper.

\section*{Competing Interests}
The authors declare no competing interests.

\newpage
\clearpage
\renewcommand\thefigure{S\arabic{figure}}    
\setcounter{figure}{0} 
\renewcommand{\theequation}{S\arabic{equation}}
\setcounter{equation}{0}
\begin{center}
	\textbf{\large Supplementary Information for "Fibonacci-Modulation-Induced Multiple Topological Anderson Insulators"}
\end{center}

This Supplementary Information details the procedures for generating generalized $\kappa$-Fibonacci-modulated chain, presents the derivation of topological phase boundaries, analyzes off-diagonal Fibonacci-modulated SOC chains, and discusses the impact of modulation frequency on topological properties. 

\subsection*{\label{sec:level1} Supplementary Note 1: Generate a generalized $\kappa$-Fibonacci-modulated chain }
Three methods are commonly employed to generate a $\kappa$-Fibonacci-modulated chain: the substitution method, the cut-and-project method, and the characteristic function method. In the substitution method, the generalized $\kappa$-Fibonacci substitution rule $\tilde{\sigma}$ acts on the two letters $A$ and $B$ as follows:
\begin{align}
	\tilde{\sigma}:\begin{cases} A \to A^{\kappa}B, \\
		B \to A,
	\end{cases}
\end{align}
where $\kappa=1,2,3,\cdots$ corresponds to different metallic mean sequences. By repeatedly applying this substitution on the initial letter $B$, one generates a sequence of words $C_l=\tilde{\sigma}^{l}(B)$ of increasing length. These finite sequences serve as approximates to the infinite generalized $\kappa$-Fibonacci chain, which is recovered in the limit $l\to \infty$. The inflation matrix associated with this substitution has eigenvalues that satisfy the characteristic equation $\lambda^{\prime 2}-\kappa \lambda^{\prime}-1=0$. The Perron-Frobenius eigenvalues for several cases are: $\lambda^{\prime}_{g}=(\sqrt{5}+1)/2$ for $\kappa=1$ (golden mean), $\lambda^{\prime}_{s}=\sqrt{2}+1$ for $\kappa=2$ (silver mean), and $\lambda^{\prime}_{b}=(\sqrt{13}+3)/2$ for $\kappa=3$ (bronze mean). The modulation frequency $\alpha$ of the Fibonacci-modulated chain is chosen as the inverse of the Perron-Frobenius eigenvalue, i.e., $\alpha=1/\lambda^{\prime}$. For numerical calculations, $\alpha$ is approximated by the rational number $\alpha=F_{\nu-1}/F_{\nu}$, where the generalized Fibonacci numbers $F_{\nu}$ are defined recursively by $F_{\nu+1}=F_{\nu-1}+\kappa F_{v}$ with initial conditions $F_0=F_1=1$.

\begin{figure*}[h]
	\centering
	\includegraphics[clip, width=0.6\columnwidth]{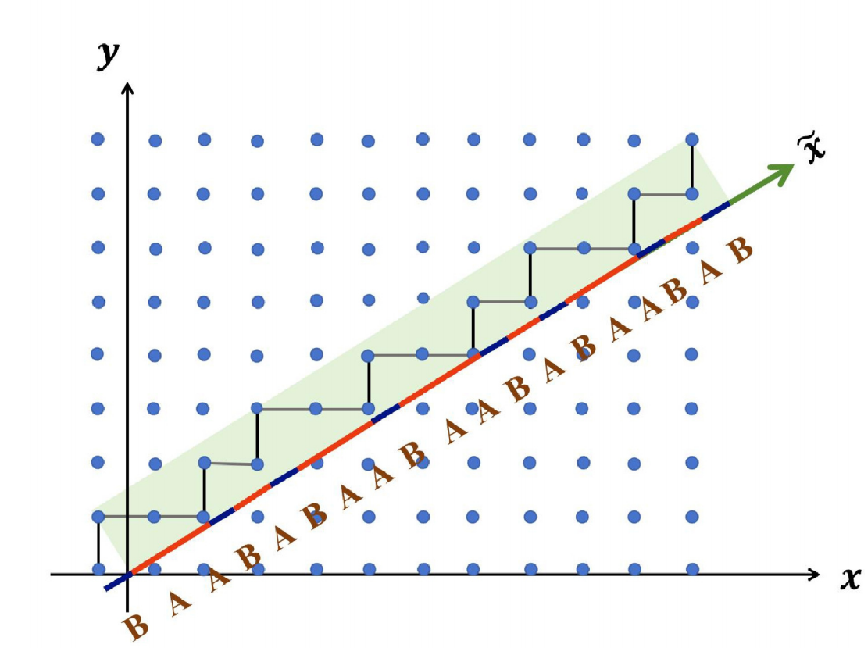}
	\caption{Schema of the cut-and project method. Selected points of a 2D square lattice are projected onto the physical axis $\tilde{x}$, giving the quasiperiodic sequence of A(red) and B(blue) tiles.}
	\label{aFig1}
\end{figure*}

The generalized $\kappa$-Fibonacci chain can also be constructed using the cut-and-project method, which projects a 2D periodic lattice onto a 1D quasicrystal. As illustrated in Fig. \ref{aFig1}, the parent system is a 2D square lattice, and the quasicrystalline chain is obtained by projecting lattice points onto the physical axis $\tilde{x}$, which is oriented at an angle $\tilde{\theta}$ such that $\tan{\tilde{\theta}}=\alpha$. A periodic lattice point is selected if it lies within a green strip of width equal to one unit cell of the 2D lattice. Upon projection onto the $\tilde{x}$ axis, the horizontal and vertical bonds of the square lattice project onto the red and blue intervals, respectively. The resulting spacings between adjacent points on the projected chain---represented by these red and blue intervals---can take two values, $\cos{\tilde{\theta}}=1/\sqrt{1+\alpha^2}$ and $\sin{\tilde{\theta}}=\alpha/\sqrt{1+\alpha^2}$. These two lengths corresponds to the $A$ and $B$ tiles in the generalized $\kappa$-Fibonacci sequence, respectively.

The third approach for generating the generalized $\kappa$-Fibonacci quasicrystal is the characteristic function method. Kraus and Zilberberg \cite{Zilberberg2012fulu} proposed a scheme that bridges the Fibonacci and AA models. In this method, the $i$-th letter of the $\kappa$-Fibonacci chain is determined by the characteristic function $\chi_i$, defined by
\begin{align}
	\chi_i=\mathrm{sgn}[\cos(2\pi \alpha i+\phi)-\cos(\pi\alpha)],	
\end{align}
where $\phi$ is an arbitrary phase. According to this definition, $\chi_i=+1$ corresponds to the letter $A$, while $\chi_i=-1$ corresponds to the letter $B$.

\subsection*{\label{sec:level2} Supplementary Note 2: Derivation of topological phase boundaries}

In the topologically nontrivial regime, the zero-energy edge modes emerge with a finite localization length. Conversely, in the trivial regime, these edge modes vanish, giving way to bulk states characterized by a divergence in localization length. To analytically determine the topological phase boundaries, we examine the localization length of the zero modes. The zero mode Schr\"{o}dinger equation for the Fibonacci-modulated model, $H|\Psi\rangle=0$, is expressed as:
\begin{align}
	-t_0(\psi_{i+1\uparrow}+\psi_{i-1\uparrow})+t_{so}(\psi_{i+1\downarrow}-\psi_{i-1\downarrow})+U_i\psi_{i\uparrow} &=0, \\
	t_0(\psi_{i+1\downarrow}+\psi_{i-1\downarrow})+t_{so}(\psi_{i-1\uparrow}-\psi_{i+1\uparrow})-U_i\psi_{i\downarrow} &=0,
\end{align}
where $\psi_{i\beta}$ represents the amplitude of the zero mode with spin $\beta=\{\uparrow,\downarrow\}$ at site $i$, and $U_i=[\lambda\mathrm{sgn}(\cos{(2\pi\alpha i+\phi)}-\cos{(\pi\alpha)})+M]$. By introducing the local basis transformation $\phi^{+}_{i}=\psi_{i\uparrow}+\psi_{i\downarrow}$ and $\phi^{-}_{i}=\psi_{i\uparrow}-\psi_{i\downarrow}$, the coupled equations decouple into two independent equations of similar form:
\begin{align}
	-t_0(\phi^{-}_{i+1}+\phi^{-}_{i-1})+t_{so}(\phi^{-}_{i-1}-\phi^{-}_{i+1})+U_i\phi^{-}_{i}&=0, \\
	-t_0(\phi^{+}_{i+1}+\phi^{+}_{i-1})+t_{so}(\phi^{+}_{i+1}-\phi^{+}_{i-1})+U_i\phi^{+}_{i}&=0.
\end{align}
Focusing on one of these decoupled equations, the evolution can be rewritten in terms of a transfer matrix:
\begin{equation}
	\begin{pmatrix}
		\phi^{-}_{i+1} \\
		\phi^{-}_{i}\end{pmatrix}=T_i\begin{pmatrix}
		\phi^{-}_{i} \\
		\phi^{-}_{i-1} \end{pmatrix}
\end{equation}
where the transfer matrix $T_i$ is
\begin{equation}
	T_i=\begin{pmatrix}	\frac{U_i}{t_0+t_{so}} & \frac{t_{so}-t_0}{t_0+t_{so}} \\ 1 & 0 \end{pmatrix}.
\end{equation}
The localization length $\tilde{\lambda}$ of the zero modes is then determined by
\begin{equation}
	\tilde{\lambda}^{-1}=\lim_{L\to\infty}\frac{1}{L}\ln{||T||},
\end{equation}
where $T\equiv\prod_{i=1}^{L}T_i$ is the total transfer matrix. Here, $\|\cdot\|$ denotes the matrix Euclidean norm, which is defined as the largest singular value of the matrix. For a normal matrix, such as the transfer matrix considered here, the Euclidean norm is equivalent to the absolute value of its largest eigenvalue. For the $2\times 2$ matrix $T$, the eigenvalues are
	\begin{align}
		\lambda_1=\frac{\mathrm{Tr}(T)}{2}+\sqrt{\left[\frac{\mathrm{Tr}(T)}{2}\right]^2-\mathrm{Det}(T)},\quad \lambda_2=\frac{\mathrm{Tr}(T)}{2}-\sqrt{\left[\frac{\mathrm{Tr}(T)}{2}\right]^2-\mathrm{Det}(T)}.
	\end{align}
The phase boundary between topological and trivial regimes is defined by the condition $\tilde{\lambda}^{-1}=0$, which requires that the largest eigenvalue of $T$ satisfies $\left|\max\{\lambda_1,\lambda_2\}\right|=1$. Solving this condition yields the phase boundary equation: $|\mathrm{Tr}(T)|-\mathrm{Det}(T)=1$, which is shown in the main text.

\subsection*{\label{sec:level3} Supplementary Note 3: Off-diagonal-Fibonacci-modulated SOC chain}
\begin{figure*}[htbp]
	\centering
	\includegraphics[clip, width=0.5\columnwidth]{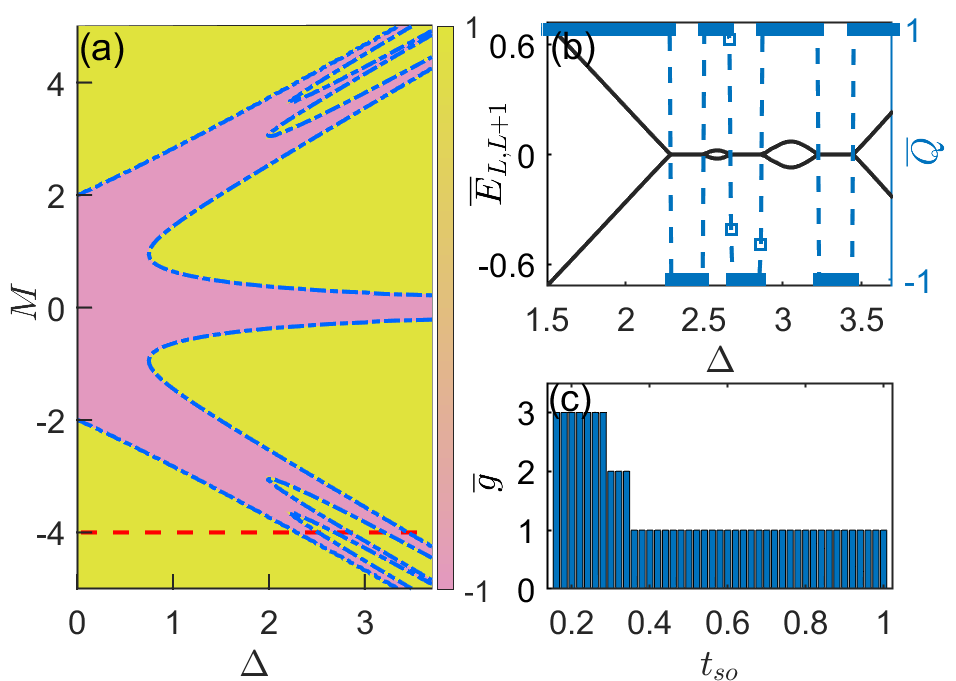}
	\caption{(a)Topological phase diagram of an off-diagonal-Fibonacci-modulated SOC chain in $\Delta-M$ plane with $L=610$, $\alpha=F_{13}/F_{14}$, and $t_{so}=0.25$. The blue dashed lines represent the topological phase boundaries numerically obtained by Eq. (3) in the main text. (b) Two disorder-averaged energies $\overline{E}_L$ and $\overline{E}_{L+1}$ at the center of the spectrum and the disorder-averaged topological number $\overline{Q}$ as a function of $\Delta$ under OBC with $t_{so}=0.25$ and $M=-4$. (c) The number of times that the TAI phase emerges as a function of $t_{so}$ with $M=-4$. Here, all the data are averaged by $N_c=50$ disorder realizations.}
	\label{aFig2}
\end{figure*}

In this section, we examine an off-diagonal Fibonacci-modulated SOC chain, described by the Hamiltonian (1) in the main text with $\mathcal{U}_i=M\sigma_z$ and
\begin{equation}
	\mathcal{R}_{ij}=-\left[t_0-\frac{\Delta}{2}\mathrm{sgn}\left[\cos(2\pi\alpha i+\phi)-\cos(\pi\alpha)\right]+\frac{\Delta}{2}\right]\sigma_z\pm it_{so}\sigma_y,
\end{equation}
where the $\pm$ corresponds to the hopping along the $\pm \hat{x}$ direction. The diagonal terms of $\mathcal{R}$ describe the spin-conserved hopping with the amplitude $t_0$, modulated by the Fibonacci sequence with strength $\Delta$. The off-diagonal terms represent spin-flip hopping between nearest neighbors with an amplitude $\mathcal{R}$. We set $t_0=1$ as the unit of energy throughout this discussion.
We calculated the topological phase diagram in the $\Delta-M$ plane, as shown in Fig. \ref{aFig2}(a), using $L=610$, $t_{so}=0.25$, and $N_c=50$ disorder realizations. The dashed lines represent the topological phase boundaries, numerically determined by the condition $|\mathrm{Tr}(T)|-\mathrm{Det}(T)=1$, where $T=\prod_{i=1}^{L}T_i$ and
\begin{equation}
	T_i=\begin{pmatrix}	\frac{M}{\xi_{i}^{(+)}} & \frac{\xi_{i-1}^{(-)}}{\xi_{i}^{(+)}} \\ 1 & 0 \end{pmatrix}.
\end{equation}
\begin{figure*}[htbp]
	\centering
	\includegraphics[clip, width=0.5\columnwidth]{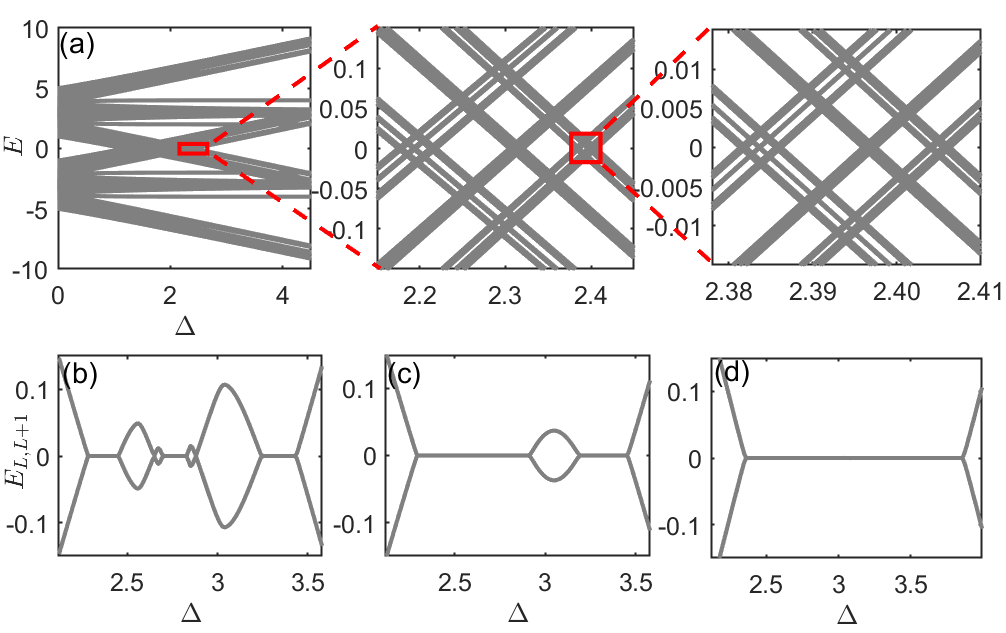}
	\caption{(a)Self-similarities of energy spectrum as a function of $\Delta$ with $t_{so}=0$. Two central energy levels $E_{L}$ and $E_{L+1}$ as a function of $\Delta$ with (b) $t_{so}=0.2$, (c) $0.3$, and (d) $0.9$.  Here, $M=-4$, $\phi=0$, and $L=610$.}
	\label{aFig3}
\end{figure*}
\begin{figure*}[htbp]
	\centering
	\includegraphics[clip, width=0.5\columnwidth]{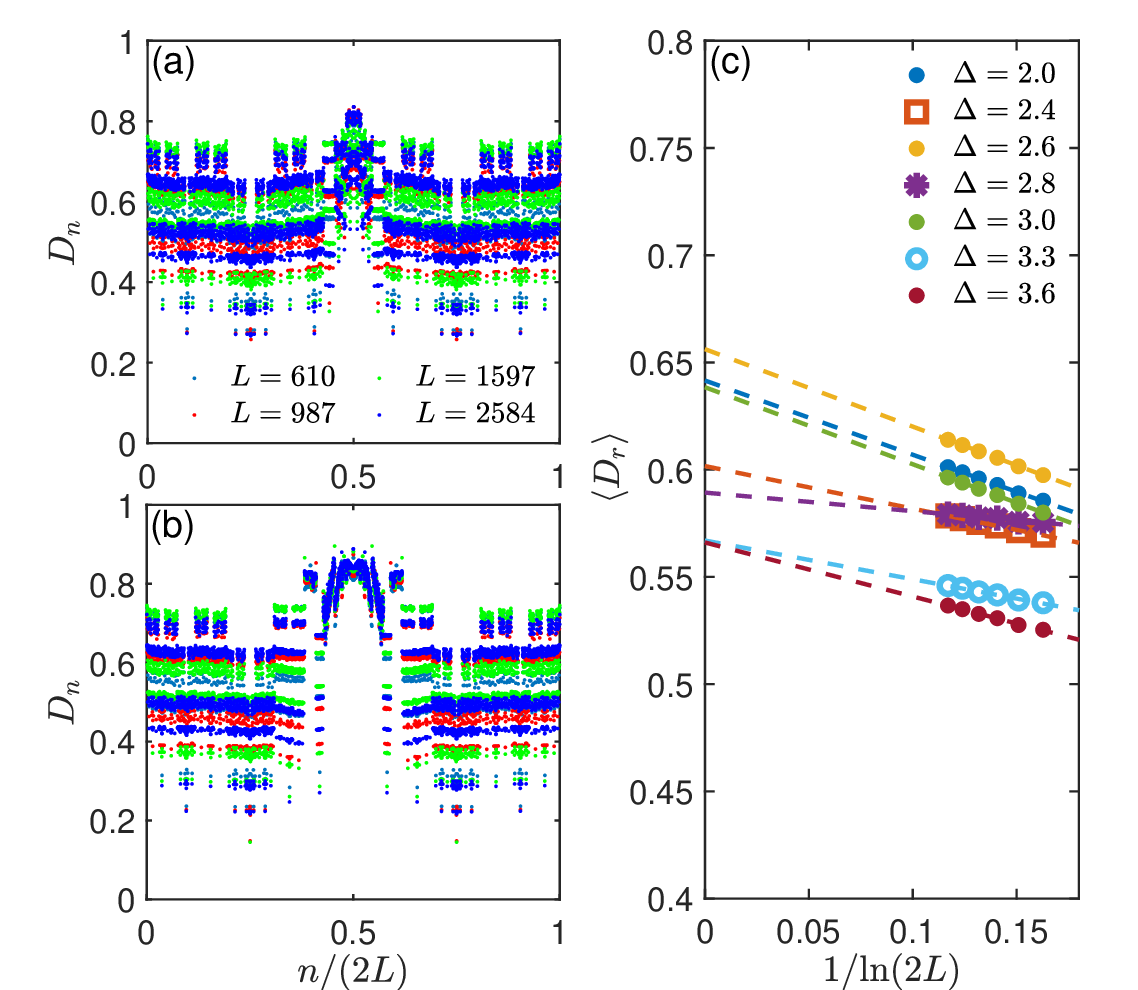}
	\caption{Fractal dimensions $D_n$ for different system sizes under PBC for (a) $\Delta=2.4$ and (b) $\Delta=3$, respectively. (c) The scaling of the mean fractal dimensions $\langle D\rangle$ for various $\Delta$ values. Here, $t_{so}=0.25$, $M=-4$, and $\phi=0$.}
	\label{aFig4}
\end{figure*}
\noindent where, $\xi_{i}^{(\pm)}=t_{so}\pm
\left[t_0-\frac{\Delta}{2}\mathrm{sgn}\left[\cos(2\pi\alpha i+\phi)-\cos(\pi\alpha)\right]+\frac{\Delta}{2}\right]$. The results show that the topologically nontrivial phase is robust against weak modulation for $0<\Delta<2$ but transitions to a trivial phase as the modulation strength increases. Notably, when $M=0$, the system remains in the topological phase regardless of $\Delta$. For $2<|M|<3.05$, the system undergoes a topological phase transition from trivial to nontrivial and back to trivial with increasing $\Delta$, demonstrating the emergence of the TAI phase. When $|M|\ge 3.05$, the system exhibits multiple transitions into the TAI phase before ultimately reverting to a trivial phase at higher $\Delta$. Figure \ref{aFig2}(b) shows the disorder-averaged energies of the two center modes, $\overline{E}_{L}$ and $\overline{E}_{L+1}$, alongside the $\mathcal{Z}_2$ topological number $\overline{Q}$ as a function of the modulation strength $\Delta$, with $M=-4$ and $t_{so}=0.25$. The results indicate that the moderate modulation $\Delta$ induces changes in the nontrivial invariant within the regions $\Delta\in[2.28,2.49]\cup[2.66,2.86]\cup[3.22,3.44]$, where disorder-averaged zero modes are also observed. For $\Delta>3.44$, the TAI phase vanishes. In Fig. \ref{aFig2}(c), we show the number of times the TAI phase emerges, $\overline{g}$, as a function of $t_{so}$ for $M=-4$. Similar to the on-site modulated case, $\overline{g}$ in the off-diagonal-modulated case increases as $t_{so}$ decreases. To understand this phenomenon, we numerically calculate the energy spectrum of two decoupled off-diagonal-Fibonacci-modulated chains without the SOC term ($t_{so}=0$) as a function of $\Delta$, shown in Fig. \ref{aFig3}(a) for $M=-4$ and $\phi=0$. Within the range $\Delta \in (1.16, 2.41)$, the middle bands overlap, forming three distinct clusters of band-crossing regions separated by two band gaps near zero energy. These clusters further split into sub-clusters, creasing a self-similar fractal structure in the thermodynamic limit, as seen in Fig. \ref{aFig3}(a). As the SOC strength $t_{so}$ increases, these band gaps, with magnitudes comparable to $t_{so}$, progressively close and reopen, leading to the emergence of nontrivial zero-energy modes in these gaps, as depicted in Fig. \ref{aFig3}(b). This suggests an infinite number of TAI phases emerge in the small $t_{so}$ limit and in the thermodynamic limit, forming a topologically fractal structure. As $t_{so}$ increases further, adjacent topologically nontrivial regions merge, reducing the number of distinct TAI phases, as illustrated in Figs. \ref{aFig3}(c) and \ref{aFig3}(d). These findings confirm that an off-diagonal Fibonacci-modulated SOC chain supports multiple TAI phases, with their occurrence strongly dependent on $t_{so}$. 

Figures \ref{aFig4}(a) and \ref{aFig4}(b) show the fractal dimensions for different system sizes under periodic boundary condition (PBC) with $\phi=0$ for $\Delta=2.4$ and $\Delta=3$, respectively. For both topologically trivial and nontrivial phases, the fractal dimensions of nearly all states are system-size independent, deviating from the extremes of $0$ and $1$. The scaling behavior of the mean fractal dimension $\langle D_r \rangle$ for various $\Delta$ values, in both topologically trivial ($\Delta=2.4$, $2.8$, and $3.3$) and nontrivial ($\Delta=2$, $2.6$, $3$, and $3.6$) phases, is shown in Fig. \ref{aFig4}(c). In the thermodynamic limit, $\langle D_r \rangle$ approaches values that deviate from both $0$ and $1$ in both phases. Our results suggest that the emergence of the TAI phase does not affect the localization properties of the system, and the states remain multifractal.

\subsection*{\label{sec:level4} Supplementary Note 4: Influence of modulation frequency on the emergence of multiple reentrant TAI phases}
\begin{figure*}[htbp]
	\centering
	\includegraphics[clip, width=1\columnwidth]{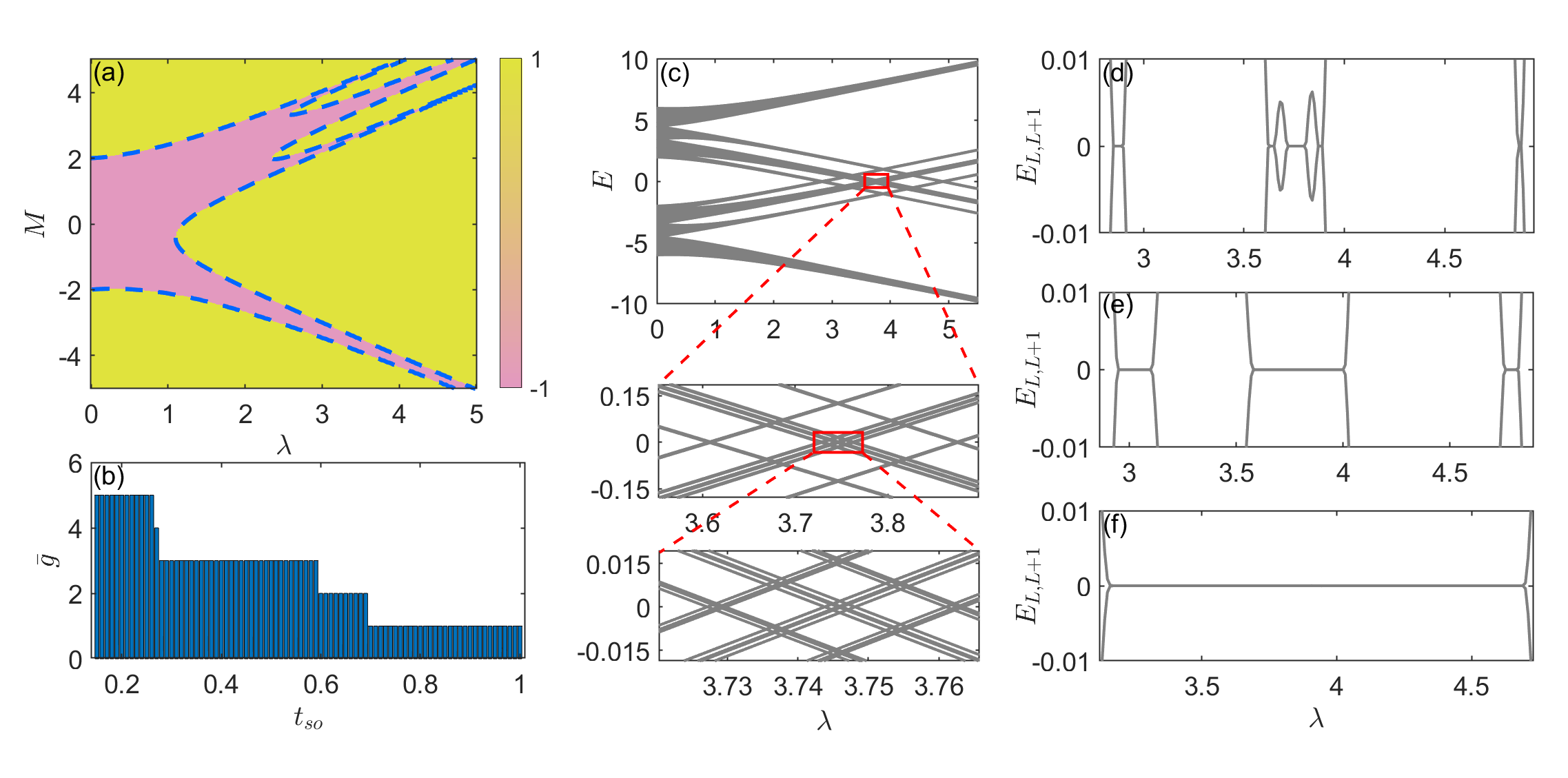}
	\caption{(a) The disorder-averaged topological phase diagram of the system in $M-\lambda$ plane with $t_{so}=0.5$. The blue dashed lines correspond to the topological phase boundaries determined by Eq.(3) in the main text. (b) The number of times the TAI phases emerge, denoted as $\overline{g}$, as a function of $t_{so}$ with $M=4$. Here, all the data are averaged by $N_c=50$ disorder realizations. (c) Self-similarities of energy spectrum as a function of $\lambda$ with $t_{so}=0$. Two central energy levels $E_{L}$ and $E_{L+1}$ as a function of $\lambda$ with (d) $t_{so}=0.25$, (e) $0.5$, and (f) $0.9$. Here, $\kappa=2$, $M=4$, $\phi=0$, and $L=577$.}
	\label{aFig5}
\end{figure*}
\begin{figure*}[htbp]
	\centering
	\includegraphics[clip, width=1\columnwidth]{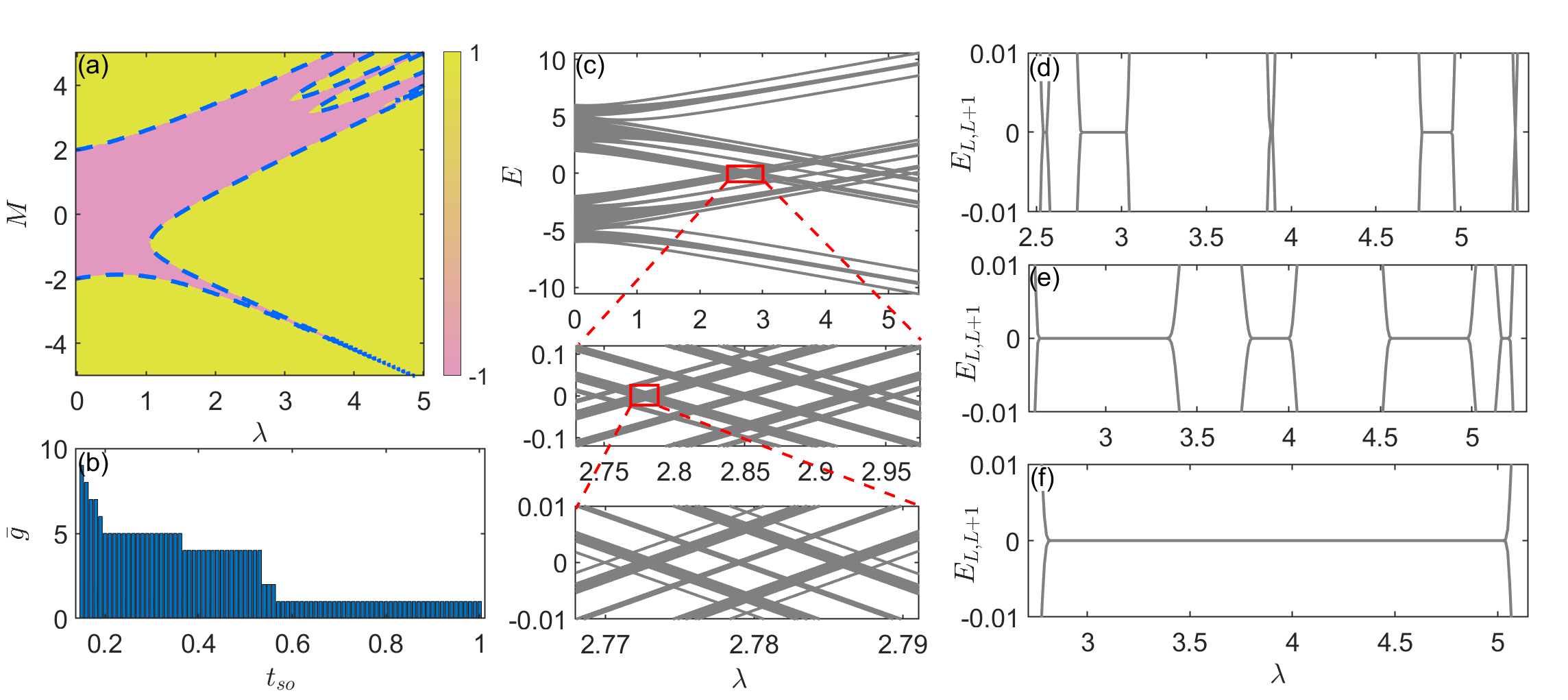}
	\caption{(a) The disorder-averaged topological phase diagram of the system in $M-\lambda$ plane with $t_{so}=0.5$. The blue dashed lines correspond to the topological phase boundaries determined by Eq.(3) in the main text. (b) The number of times the TAI phases emerge, denoted as $\overline{g}$, as a function of $t_{so}$ with $M=4$. Here, all the data are averaged by $N_c=50$ disorder realizations. (c) Self-similarities of energy spectrum as a function of $\lambda$ with $t_{so}=0$. Two central energy levels $E_{L}$ and $E_{L+1}$ as a function of $\lambda$ with (d) $t_{so}=0.27$, (e) $0.5$, and (f) $0.9$. Here, $\kappa=3$, $M=4$, $\phi=0$, and $L=469$.}
	\label{aFig6}
\end{figure*}

To investigate the influence of modulation frequency on the emergence of multiple TAIs, we extend our analysis to other metallic mean numbers ($\kappa=2$ and $3$), in contrast to the main text where $\alpha$ is chosen as the golden mean $(\sqrt{5}-1)/2$, corresponding to $\kappa=1$.

For the silver mean ($\kappa=2$), Fig. \ref{aFig5}(a) displays the disorder-averaged topological phase diagram for $t_{so}=0.5$ as a function of $M$ and $\lambda$. This diagram is obtained by numerically averaging $\mathcal{Z}_2$ invariant $\overline{Q}$ over $N_c$ disorder realizations for various values of $\phi\in[0,2\pi)$, with phase boundaries determined according to Eq.(3) in the main text. For $M>2$, the system exhibits multiple distinct transitions into the TAI phase, eventually reverting to a trivial phase as $\lambda$ increases further. We also examine how the number of TAI phases varies with $t_{so}$ at $M=4$, as shown in Fig. \ref{aFig5}(b). It is evident that the average number of emergent TAI phases $\overline{g}$ increases as $t_{so}$ decreases. Following the approach in the main text, we further analyze the energy spectrum of two decoupled Fibonacci-modulated chains with $\kappa=2$ in the absence of the SOC term ($t_{so}=0$), as depicted in Fig. \ref{aFig5}(c) for $\phi=0$. Here, a self-similar fractal structure is observed. Upon introducing a finite SOC strength $t_{so}$, band gaps comparable in size to $t_{so}$ begin to close and reopen, indicating the emergence of nontrivial zero-energy modes, as illustrated in Fig. \ref{aFig5}(d). As $t_{so}$ increases further, adjacent topologically nontrivial regions merge, reducing the number of distinct TAI phases. This transition, shown in Figs. \ref{aFig5}(e) and \ref{aFig5}(f), marks the evolution from a fractal topological structure to a single, more conventional one.

For the bronze mean ($\kappa=3$), Fig. \ref{aFig6}(a) shows the disorder-averaged topological phase diagram for $t_{so}=0.5$ as a function of $M$ and $\lambda$, obtained using the same methodology. The phase boundaries are again determined by Eq.(3) in the main text. For $M>3.1$, the system undergoes multiple distinct transitions into the TAI phase, eventually returning to a trivial phase as $\lambda$ increases further. Figure \ref{aFig6}(b) presents the dependence of the number of emergent TAI phases on $t_{so}$ at $M=4$. We find that $\overline{g}$ increases as $t_{so}$ decreases, with a growth rate even faster than in the golden mean and silver mean cases. Using the same approach, we analyze the energy spectrum of two decoupled Fibonacci-modulated chains with $\kappa=3$ without the SOC term ($t_{so}=0$), as shown in Fig. \ref{aFig6}(c) for $\phi=0$, and again observe a self-similar fractal structure. When the SOC strength $t_{so}$ is introduced, band gaps comparable in size to $t_{so}$ start to close and reopen, signaling the formation of nontrivial zero-energy modes, as depicted in Fig. \ref{aFig6}(d). As $t_{so}$ increases further, neighboring topologically nontrivial regions merge, reducing the number of distinct TAI phases. This transition, illustrated in Figs. \ref{aFig6}(e) and \ref{aFig6}(f), signifies a transformation from a fractal topology to a more discrete structure.

Overall, our findings demonstrate that the emergence of multiple reentrant TAI phases is a universal and robust phenomenon across different types of Fibonacci modulation frequencies, consistent with the results presented in the main text. By employing various metallic mean modulations, such as the silver and bronze means, we consistently observe self-similar fractal structures in the energy spectra and multiple transitions between trivial and topological phases as the disorder strength varies. Notably, the modulation frequency not only determines the complexity of the fractal structure but also significantly influences the evolution of TAI phases. These results highlight the universal and robust nature of reentrant TAI behavior induced by fractal modulations, and underscore the critical role of modulation frequency in shaping the topological phase diagram of the system.

\end{document}